%% file: 2573.tex
\documentclass[]{aa} %
\usepackage{graphicx}
\usepackage{txfonts}
\begin{document}

\title{ The Kyiv Meridian Axial Circle Catalogue of stars
 in fields with extragalactic radio sources
       \thanks{The catalogue is only available in electronic form
	at the CDS via anonymous ftp to cdsarc.u-strasbg.fr (130.79.128.5)
	}
}	

\author{   P.Lazorenko \inst{1}, Yu.Babenko \inst{2},
	 V.Karbovsky \inst{1}, M.Buromsky \inst{2},
         O.Denisjuk \inst{1}, \and  S.Kasjan \inst{2}}

   \offprints{P.Lazorenko}

     \institute{ Main Astronomical Observatory,
          	National Academy of Sciencies of Ukraine,
             Zabolotnogo 27, 03680 Kyiv-127, Ukraine\\
             email: laz@mao.kiev.ua 
            \and   
		Astronomical Observatory of the Kyiv
		National University, Observatornaya 3,
		 04053  Kyiv-53, Ukraine\\
		email: babenko@observ.univ.kiev.ua       
	}

   \date{Received December 20, 2004; accepted March 07, 2005}

   \abstract{ 
A catalogue of astrometric (positions, proper motions) and photometric 
(B, V, R, r$'$, J) data 
of stars in fields with ICRF objects has been compiled  at the Observatory 
of the 
National Academy 
of Sciences of Ukraine and the Kyiv University Observatory. All fields are 
located in the declination 
zone from 0$^{\circ}$ to +30$^{\circ}$; 
the nominal field size is 46$'$(right ascension)x24$'$ 
(declination). The 
observational basis of this work is 1100 CCD scans down to V=17 mag which were 
obtained with the Kyiv meridian axial circle in 2001--2003.
The catalogue is presented in two versions. The 
version KMAC1-T contains 159 fields (104\,796 stars) and was obtained with 
reduction to the 
Tycho2 catalogue. For another 33 fields, due to a low sky density of 
 Tycho2 stars, the reduction 
was found to be unreliable. Transformation to the ICRF
system in the second version of the catalogue (KMAC1-CU) was 
performed
using the UCAC2 and CMC13 catalogues as a reference; it contains 115\,032 stars 
in  192 fields and is of slightly  better accuracy.  The
external accuracy of one catalogue position is about 50--90~mas 
for V$<$15~mag stars. The average error of photometry is better than 
0.1~mag for stars down to 16~mag.


}

  \titlerunning{ The Kyiv Meridian Axial Circle Catalogue}
	\authorrunning{P.Lazorenko  et al.}

    \maketitle

\input 2573_1

\input 2573_2

\input 2573_3
\input 2573_4
\input 2573_5
\input 2573_6

\begin{acknowledgements}
 We  acknowledge Dr. D.W.Evans important observation concerning
photometric calibrations, 
V.Andruk for his suggestions  on the CCD raw data filtering 
and image processing, and
Dr. A.Yatsenko for valuable remarks about proper motion determination.
    Technical development of the CCD micrometer was carried out by
O.Kovalchuk (Nikolaev Astronomical Observatory, Ukraine).
    This publication makes use of data products from the Two Micron All
    Sky Survey, which is a joint project of the University of Massachusetts
    and the Infrared Processing and Analysis Center/California Institute of
    Technology, funded by the National Aeronautics and Space Administration
    and the National Science Foundation; the VIZIER database, operated at CDS,
Strasbourg, France; the WEBDA database for open cluster; 
and NASA's Astrophysical Data System Abstract Service.

\end{acknowledgements}

\input ref.tex
\end{document}

%% file: 2573_1.tex
\section{ Introduction}
The Meridian Axial Circle (MAC, D=180~mm, F=2.3~m) in Kiev was recently 
modernized by installing a 1040x1160 CCD camera that can work
in scan mode (Telnyuk-Adamchuk et al. \cite{aa2002}; Karbovsky \cite{karb}).
The camera, designed at the Nikolaev Observatory (Ukraine),
incorporates a glass  filter  to enable observations
in the V band. With effective exposures of
about 108~sec for equatorial stars, the magnitude limit is V=17~mag. 

	The instrument was used in  two observational projects.
The first long-term  project was the astrometric survey of the sky in 
the equatorial 
zone to  extend the Hipparcos-Tycho 
reference frame to fainter magnitudes.  This programme is still in progress.
The second project, now completed,  concerns observations of star fields 
in the direction of 
192 extragalactic ISRF objects, a list of which, for the declination zone 
from 0$^{\circ}$ to +30$^{\circ}$, was taken from Molotaj (\cite{Molotaj}).
This declination range was chosen to reduce  CCD distortion effects
(Vertypolokh et al. \cite{nik2001}; Vertypolokh et al. \cite{journ}).
         
The project was carried out in the framework of scientific problems: 
maintenance of the Hipparcos 
frame of reference and the linking of optical frames to the ICRF. 
This report describes the data reduction
and compilation of the Kyiv meridian axial circle catalogue  (KMAC1) of stars 
in fields of   extragalactic radio reference frame sources.

The most important data sources used for the compilation of the
catalogue include the major catalogues
Tycho2 (Hog et al. \cite{tycho}); CMC13 (Evans et al. \cite{evans}); 
UCAC2 (Zacharias et al. \cite{ucac}); 2MASS (Cutri et al. \cite{2mass}); 
USNO-A2.0 (Monet et al. \cite{a20}) and USNO-B1.0  (Monet et al. \cite{b10}). 
Also, for
calibration of the instrumental magnitude scale we used several 
photometric catalogues of  NGC 2264 stars.

\begin{figure}[htb]
\centerline{\includegraphics*[width=9.0cm]{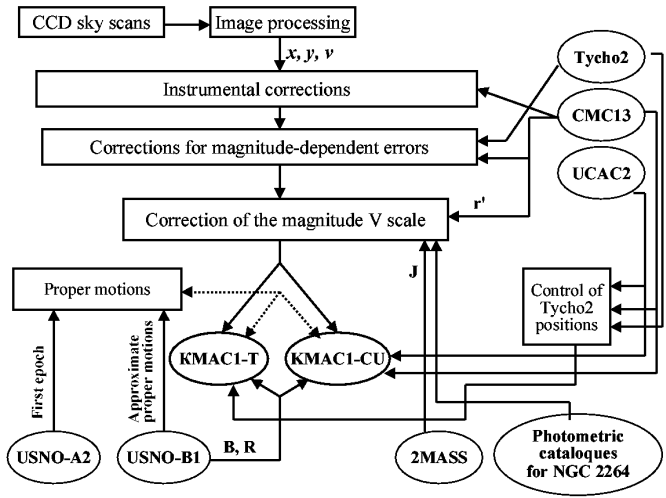}}
\caption{Compilation of the KMAC1: main steps of reduction
and source catalogues}
\label{general}
\end{figure}

The astrometric reduction and source catalogues
used for compilation of the KMAC1 are shown in Fig.~\ref{general}. 
Compilation of 
the catalogue followed the following steps of   data reduction:
image processing (Sect.~2), 
calibration for instrumental and magnitude-dependent errors (Sect.~3)
and correction of the magnitude scale (Sect.~4).
Conversion to the ICRF was carried out with the two alternative 
types of referencing,
using the space-based catalogue Tycho2 and the modern ground-based catalogues
CMC13 and UCAC2. This resulted   
in the compilation of two catalogue versions: KMAC1-T and KMAC1-CU. 
The details of
referencing to the ICRF system are discussed in Sect.~5 and
the computation of proper motions in Sect.~6.
The catalogue description, its 
properties and external verification are described in Sect.~7.

%%%%%%%%%%%%%%%%%%%%%%%%%%%%%%%%%%%%%%%%%%
%[angle=90, width=8.0cm, height=6cm]{fg1.ps}}
%%%%%%%%%%%%%%%%%%%%%%%%%%%%%%%%%%%%%%%%%%

%% file: 2573_2.tex
\section{ Image processing}
 The  catalogue is based on 1100 CCD scans each of 
46x24$'$ size in the sky (right ascension x declination) 
and centered on the observed ICRF object with
an accuracy of about $\pm 2'$. Each of the 192 ICRF fields 
was scanned on at least  5 nights. The original 
scanned data were archived and stored in a CD-ROM database.

The first stage of data reduction began with a
search and extraction of data files  from the database 
archive.  CCD images of stellar fields were then filtered of various 
instrumental and noise
features that  introduce an inhomogeneity in the sky level. 
The inhomogeneity pattern inherent to  a scan mode is dominated by a 
1D strip-like structure that changes only along the declination 
(DEC) direction (the $x$-axis in the CCD), 
with a possible weak 
trend over right ascension (RA), the $y$-axis of the CCD. 
The striped structures in the images  are formed by increased noise from 
a few dozen  bad bright pixels, which produce vertical  pixel-width 
noisy strips. Images are also contaminated by
 a number of flares and tracks of radioactive particles of cosmic origin
and from Chernobyl and which have coma or star-like shapes.
Also, the sky level measured along the $x$-axis has a large-scale component
which under normal observing conditions does not exceed 5\% of the
total signal level. Some scans also show vertical variations
in the sky level related to clouds or changing sky brightness.

All types of background variation were eliminated with a simple correction
model that considered these variations to be  caused by additive components. 
While this interpretation is reasonable for a vertical pixel width 
structure, large-scale
variations along the $x$-axis can also contain
a multiplicative flat field component. To investigate this problem, 
we carried out a study illuminating the CCD with a light source
placed at the telescope objective. Using bias information read from the
outer calibration regions of the CCD, the flatfield pattern was computed
and compared to the systematic trends in the 
preliminary differences of instrumental
$v$ magnitudes and $r'$ CMC13 photometry. Only a partial correlation was
found, indicating a possible variation of the bias along the $x$-axis
(a similar conclusion was reached by Evans et al. (\cite{cmc}) for
observations at the Carlsberg meridian circle). 
Considering the small amplitude of variations, they were treated
as additive components.  However, possible inaccuracies 
due to omission of multiplicative components in the image
analysis is  compensated for by the 
method of calibration
for errors dependent on instrumental parameters,
in which any residual systematic trend along the $x$-axis is eliminated
using information from an external catalogue (Sect.~3.1).

Thus, scans were filtered  by, first, subtracting   
the local sky large-scale changes in the two directions,
and then subtracting a running average taken along each column of 1 pixel
width.

Detection of  objects in the noisy field was carried out
by application of a smoothing filter whose shape approximately
corresponded to the Point Spread Function,  
and by the elimination of bright 1x1~pixel flares. Detection consisted of a
comparison of the pixel flux with a threshold 
defined as $I_{det}=[1.1+(\sigma_{n}-12)/45]\sigma_{n}$, 
where $\sigma_{n} \geq 12$ is the local sky noise. The second term
in this expression ensures approximately constant, independent of
$\sigma_{n}$ and the sky star density,  the  number of false
detections 
(from 300 to 500  per  frame). For faint images, it was required that
an object should fill at least two adjacent pixels.
For bright images a special filtration was applied to avoid
false multiple image detections.

Determination of the $x$, $y$ positions and  fluxes $v$ for each object was 
performed with the various approaches available for processing of
CCD images. These are: 1) the modified Center of Gravity  (CoG)
method (Irwin \cite{irwin}) and 2) a group of the full profile fitting 
methods based on the Gaussian linearized 
least squares method (e.g. Condon \cite{condon}; Viateau \cite{vit}). 

The modified CoG method used at the CMT (Evans et al. \cite{cmc}) 
is based on theoretical considerations by Irwin (\cite{irwin}) who
demonstrated that
its accuracy is almost equal to that obtained with a full profile
fitting. 
The  method, based on  profile fitting, provids  both for  circular 
and elliptical Gaussians; in the second case, horizontal orientation
of semi-axes was considered as adequate.  
The original non-smoothed scans were used for the image processing.
Numerical procedures corrected for the undersampling effect that occurs
when  the pixel size is large and comparable to the FWHM (Viateau \cite{vit}).
In bright images, saturated pixels were not used for the fitting. 

Centroiding was performed, trying the CoG method and the
Gaussian circular and elliptic models in turn.  When a solution
was not achieved at any step of the computation, the image quality index
was flagged as non-standard centroiding. This occured also when
a final solution, with reference to the initial approximate position
(found from the first CoG iteration) was shifted by more than 1.5 pixels.
The image quality index thus marks images that are possibly multiple
or of non-standard shape.

Computations made by different methods produced very similar results,
which supports the conclusions of Irwin (\cite{irwin}). Thus, the r.m.s.
difference of coordinates computed by the CoG and Gaussian methods is
about $\pm 0.05-0.06''$ for V=15--16~mag stars and is negligibly small
in comparison to the internal random error $\pm 0.2-0.3''$ 
of one observation. 

The most important feature of the profile fitting methods is
the possibility to change their performance so as to minimize the influence 
of systematic errors typical of the CCD used at the MAC and which seriously
degrade the accuracy of the DEC measurements (see discussion in the 
next Section). 
Preliminary processing
showed that these errors appear as a systematic trend 
in declination with magnitude, which does not depend on whether computations
are made by the CoG or profile centroiding methods. The largest effect
occurs for bright magnitudes; thus for V=10~mag
stars the systematic effect, measured with reference to V=14~mag stars,
is $0.45''$. To reduce this effect, each pixel and the related
equation of the linearized system of equations was weighted by a
factor $p=\sigma _n / \sqrt{\sigma _n^2 +I}$ where $I$ is the flux received
by the pixel from the star. This  modification of the least
squares procedure decreased the amplitude of the error to $0.15''$.

%% file: 2573_3.tex
\section{Astrometric calibrations}
The main goal of the astrometric calibrations described in this Section
is the refinement of the 
measured $x$, $y$ positional and $v$ photometric data influenced by various
bias sources that are particularly intricate for the 
CCD camera used for the observations. 
A problem arose from the inaccurate tuning
of the electronics which produced a slight asymmetry of the star images.
Based on visual inspection, we considered the effect to be acceptably
small and so started observations. 
After a major part of the observations had
been obtained, it became clear that the data is affected by large systematic
errors caused, most likely, by the asymmetry of the images. 
Thus, refinement of the data required the use of 
 special data processing technique. 

\subsection{Calibration of  instrumental errors}
\begin{figure}[htb]
\resizebox{\hsize}{!}{\includegraphics{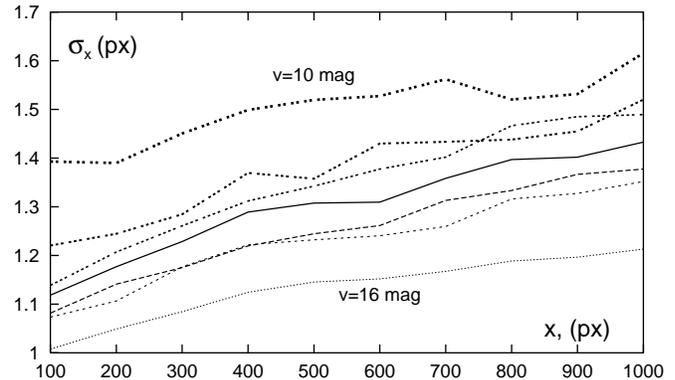}}
%\centerline{\includegraphics*[width=8.0cm]{r1.ps}}
\caption{Systematic dependence of the image size parameter 
$\sigma_x$ on the CCD $x$-coordinate. Different line types
correspond to instrumental magnitudes $v$ from 10 to 16}
\label{r1}
\end{figure}

The dominant componets of the MAC instrumental errors are related
to the following effects:  
oversaturnation of bright V$<12$~mag images caused by
use of a 12-bit AD convertor;  a slight
asymmetry of stellar profiles in the direction of the CCD 
declination $x$-axis,
along the direction of 
fast charge transfer to the reading register in the last row.
Also, profiles of star images are elongated along the 
$x$-coordinate.
While the Gaussian image size parameter $\sigma_y$ does not show any
change with $x$, the $\sigma_x$ parameter progressively increases
in this direction (Fig.~\ref{r1}), as does the image elongation. 
 Only near the CCD reading register location, 
at its left edge ($x=0$), are the images perfectly
round ($\sigma_x = \sigma_y$). 
This effect is similar to the charge
transfer efficiency problem that occurs along the scan direction
(Evans et al. \cite{cmc}), but of different origin.
No dependency of the image elongation on the background level is seen.
The amplitude of each type of image distortion
was found to depend on the star flux. Images are fairly
symmetric along the drift scan direction, so degradation due to the 
above effects
concerns mainly the  DEC and photometry.

%%%%%%%%%%%%%%%%%%%%%

%%%%%%%%%%%%%%%%%%%%
\begin{figure}[htb]
\resizebox{\hsize}{!}{\includegraphics{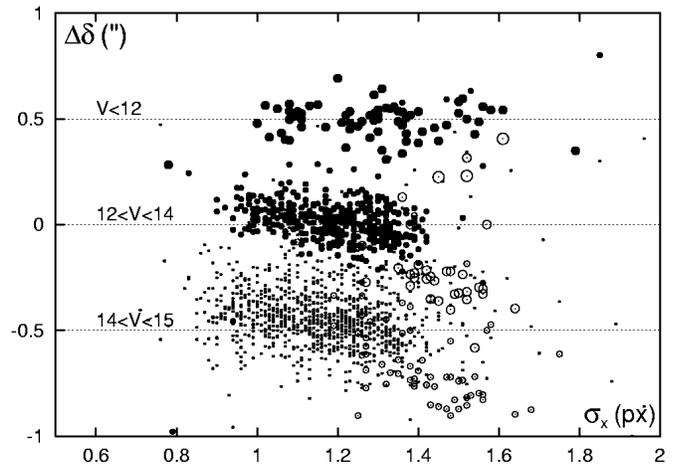}}
%\centerline{\includegraphics*[width=8.0cm]{r2.ps}}
\caption{Preliminary  differences 
KMAC1-CMC13 in DEC plotted versus $\sigma_x$ for
a few scans; stars are devided into three groups depending on $v$ and
shifted vertically by $\pm 0.5''$ for clearness. 
Open circles refer to star images with the largest $x>1050$~px
separation from the CCD reading register. Symbol size is proportional to $v$}
\label{r2}
\end{figure}
%%%%%%%%%%%%%%%%%%%%%

Image distortions, by affecting the $x$ positions of stars, 
cause a systematic bias in the DEC.
Analysis of the KMAC1 positions, obtained with  preliminary
data reduction with reference to 12--14~mag CMC13 stars, 
revealed a correlation between $\Delta \delta$ differences 
KMAC1-CMC13 and $\sigma_x$. 
Fig.~\ref{r2} shows the typical systematic trend in $\Delta \delta$, 
which is different for different magnitudes and
normally does not exceed  $\pm 0.1$--$0.2''$.
The trend is quasi-linear with a slope that depends on $v$
but that cannot be  approximated easily since a more complex
cross-relation between $\Delta \delta$, $\sigma_x$, $x$ and $v$
occures.
In particular, stars imaged in the 50~px edge area most distant from the
reading register escape this dependency.

To remove the dependence of $\Delta \delta$ on $x$ and $\sigma_x$, 
we considered a
number of models and found that the best correction is to introduce
directly to the measured $x$ values the factor:
\begin{equation}
\label{eq:r1}
\Delta x= A_v(\sigma_x - \sigma_0)
\end{equation}
where $A_v$ is a coefficient defined for each 1-mag bin of star magnitudes
and $\sigma_0$ is a constant model parameter valid for the whole data set. 
The  function
(\ref{eq:r1})  adequately models the complex nature of image distortions
inherent to the MAC, the model parameter $\sigma_0$ 
is the  $\sigma_x$ value corresponding to 
non-distorted star images, such as those  observed at the CCD reading
register ($x=0$) and are of circular form $\sigma_x=\sigma_y=\sigma_0$.
Thus  model (\ref{eq:r1})  calibrates
the $x$ coordinates for the $\sigma_x$   deviations from
$\sigma_0$, irrespective of the star position in the CCD frame and the seeing.
The use of a fixed constant $\sigma_0$ value for any magnitudes
implies that the reduction (\ref{eq:r1}) calibrates the data to a fixed star
brightness.  More complicated versions of the reduction
model that included the $x$ term or  image elongation lead to no
improvement.         

The coefficients $A_v$ and $\sigma_0$ were found
based on a criterion of best convergence of star declinations 
computed for the nights when they were observed; the 
reduction procedure used the CMC13 catalogue
as a reference. The numerical estimate of 1.11~px obtained for $\sigma_0$
corresponds to the  $\sigma_x$ value
typical for  well-exposed images of $v=13.1$~mag stars
measured near the CCD reading register ($x=0$).

A similar calibration procedure, based on a formal reduction to the 
CMC13 $r'$ photometry, was applied to instrumental magnitudes $v$. The 
difference of photometric bands is of  minor importance here
since color residuals $v-r'$ are not correlated with  the image
parameters measured at the MAC. The calibration has a form similar to
(\ref{eq:r1}):
\begin{equation}
\label{eq:r1v}
\Delta v= A'_v(\sigma_x - \sigma'_0) .
\end{equation}
Here the definition of
$\sigma'_0$ as a free model parameter lead to the appearance
of an extra systematic trend of $v$ with $x$, therefore $\sigma'_0$ was
taken to be equal to the expectation (an average) of $\sigma_x$ 
at given $x$ and $v$.
The only model parameter $A'_v$ (defined for each 1-mag bin in $v$)
was determined similarly, from a condition of the best convergence
of individual observations.

\begin{figure}[tbh]
\centerline{%
%\begin{tabular}{c@{\hspace{0.3in}}c}
\begin{tabular}{r@{}l}
\includegraphics*[height=150pt]{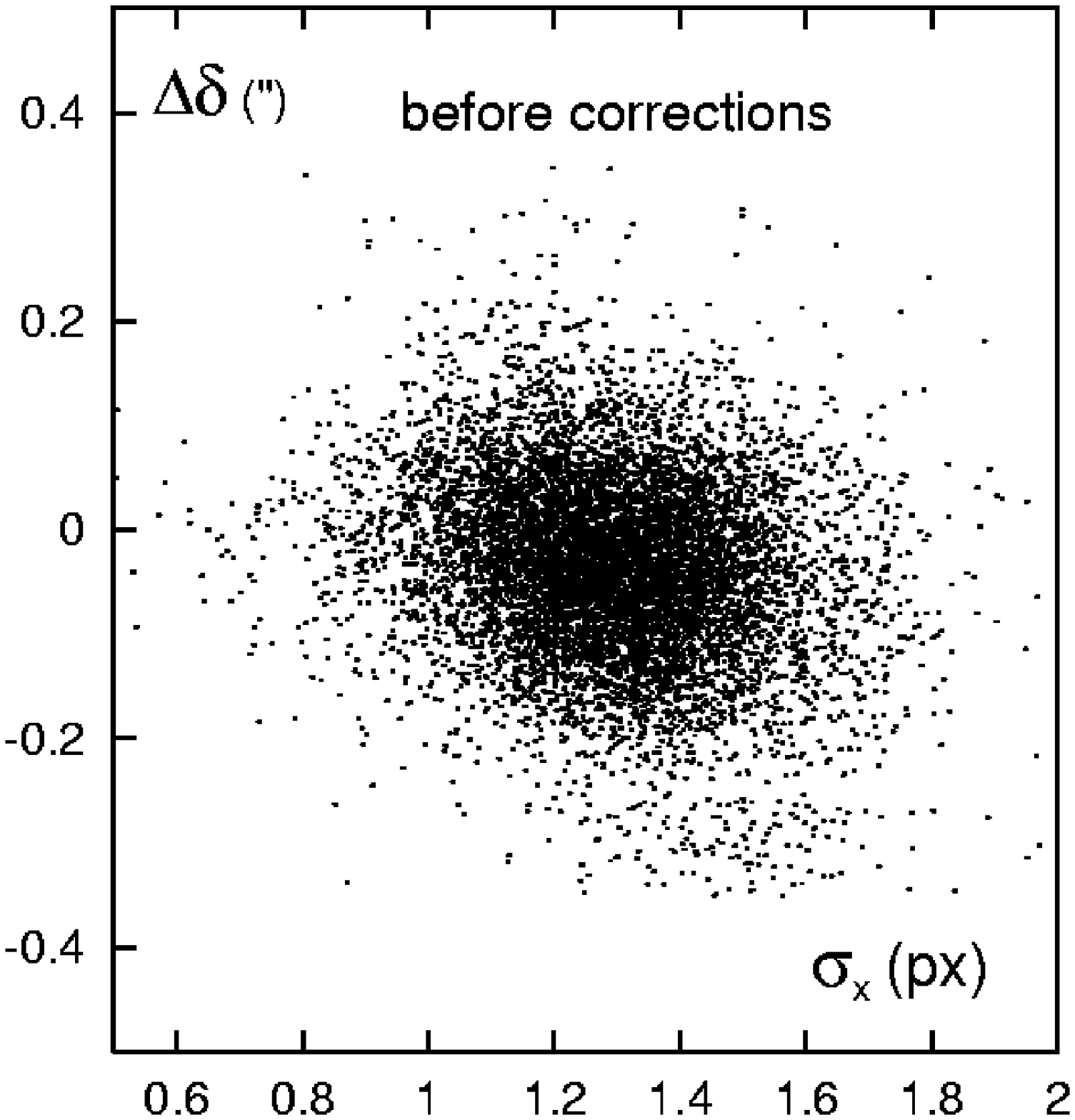} &
\includegraphics*[height=150pt]{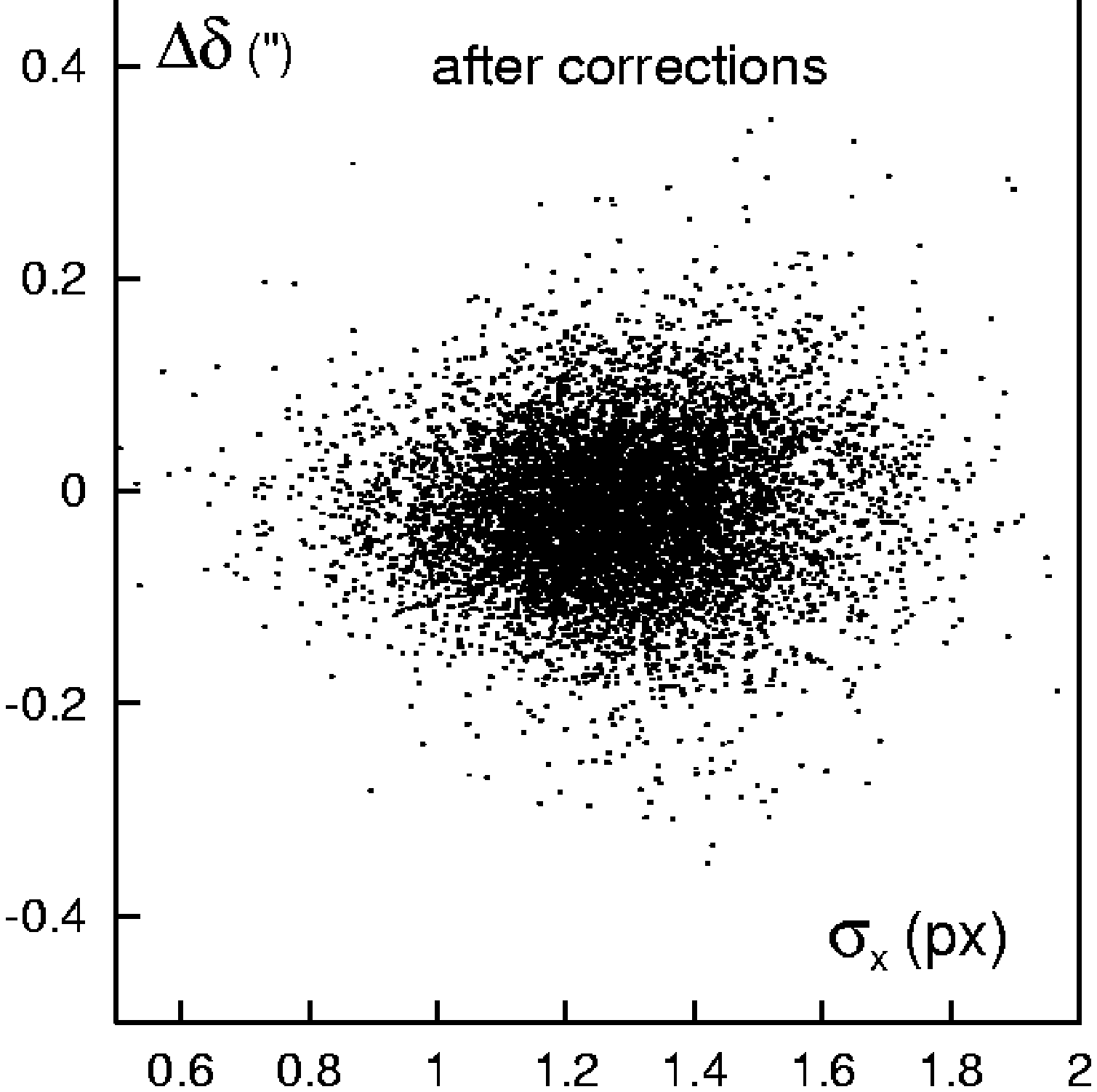} \\
\end{tabular}}
\caption{KMAC1-CMC13 differences $\Delta \delta$ versus $\sigma_x$ for
the stars of 12--14~mag: before and after calibration for 
instrumental errors}
\label{r3}
\end{figure}

Another systematic effect was found considering 
preliminary differences KMAC1-CMC13 of positions and photometric
values (formal in the last case)
computed with  calibrations (\ref{eq:r1}) and (\ref{eq:r1v}).
The differences were found to contain a small 
fluctuating component along the $x$-axis, normally
within $\pm 0.04''$ in position  and $\pm 0.03$~mag in photometry.
However in the $x >1050$~px area, the trend in DEC 
increased to $0.2''$. The origin of this trend is unclear and possibly
can be due to the imperfect pixel geometry of the CCD. 
Using  KMAC1-CMC13 differences,
the trend was removed (only the variable part, so as not to incorporate
possible systematic errors of the CMC13) and  the reduction
procedure, including determination of $A_v$, $A'_v$ and $\sigma_0$
values, was iteratively repeated. In Fig.~\ref{r3}, the 
KMAC1-CMC13 residuals in DEC
before and after calibration are shown. Along with a complete remove of
the correlation, the random scatter of $\Delta \delta$ differences 
has been noticeably reduced.

The successful refinement of the measured data suffering from
various instrumental errors was based on  extensive use of the
CMC13 as a tool for error calibration. Under the reasonable assumption of
no correlation between instrumental errors of the two telescopes,
the procedure is correct, and the accuracy of calibration
depends on the accuracy of the data in the source catalogue.
%%%%%%%%%%%%%%%%%%%%%

\subsection{Formation of equivalent scans}

In drift scan mode, formation  of star images is non-synchronous
since the median moment of the image exposure depends on RA.
For that reason, the atmospheric turbulent conditions
under which the images are formed vary as a function of RA.
The measured $x$, $y$, $v$ data are therefore  affected by a time-dependent
component of atmospheric refraction, causing an effect of
image motion much larger than  is inherent to the astrographic
mode of observations. The induced temporal signal 
is difficult to trace and
makes referencing of the observed data to the celestial system 
more difficult. A number of methods have been proposed to calibrate this 
effect, see e.g. Evans et al. (\cite{cmc}); Viateau et al. (\cite{vit}).

In the case of short scans obtained at the MAC, direct calibration of
atmospheric fluctuations with use of the Tycho2 catalogue was found to give
unreliable results since scans often contained few reference stars.
We used a method which consists of substitution of all
individual overlapping (normally to $\pm 2'$) scans available
for the particular ICRF field by a single specially-formed "equivalent" scan.
For this, each scan of the ICRF field was preprocessed 
with the 
Tycho2 catalogue so as to determine a zero point of CCD positions and 
magnitudes, and to approximately (to $\pm 1''$) reduce 
the relative displacement of individual night scans.
After cross-identification, a compiled list of field objects was formed with
$x$, $y$, $v$ data averaged. 
This procedure is similar to the formation of subcatalogues adopted
at the Valinhos meridian circle (Viateau et al. \cite{vit}) but with
no conversion to equatorial coordinates. 
The validity of this substitution is based on the linearity of the averaging
operation. Thus, the averaging can be performed either 
prior to conversion
to the celestial coordinates (that is, over the CCD measured data),
or after this reduction (over equatorial positions), with equivalent results. 
A stringent linearity of the averaging is only achieved, however, when
the star content of individual scans is identical.
This is the case for the bright stars 
that are normally detected and measured in each nightly scan.

 In the case of omitted (usually faint) object images, corrections allowing
for the compensation for the scan system of the "omitted" star observation 
should be
applied to that object's $x$, $y$, $v$ data in the equivalent scan.
The information necessary to make this correction is found by
obtaining the differences between each nightly scan and the correspondent
equivalent scan. Since we consider the measured data, not
transformed to celestial coordinates and magnitudes, the
differences usually show systematic trends in both $x$ and $y$ 
directions and due to possible varying magnitude
errors in $v$. 
The systematic component of these differences was approximated
with cubic spline functions. 

Calibrations for omitted images started from consideration of major
systematic trends along the temporal $y$-axis. After  corrections
in the equivalent scan data ($x$, $y$, $v$), 
these trends were removed
from the nightly scans. As a result, the differences between each nightly
and  "equivalent" scan  became noise-like in shape. 
Next, similar steps of calibration were applied to
the differences of "nightly scan" - "equivalent scan" registered
along the $x$ and $v$ data axes.
To obtain convergence, the whole procedure was reprocessed twice,
the outliers removed and all computations repeated again.

The resulting equivalent scans used for transformation to the ICRF
are less subject to atmospheric differential image motion due to 
averaging over a subset of individual scans included in the output.
The averaging effect is inversely proportional to the square root of the
number of frames, which is 6 on average.
The output nightly scans were
of less importance since they 
were tightly reduced to the system of the corresponding
equivalent scan by filtering out any systematic differences. 
The differences between these scans contained
only a random noise component which provided 
valuable information on internal catalogue errors (Sect.~7.1). 

An important restriction to the method discussed is
that correct tracing of scan system changes is achieved only with 
completely overlapped and co-centered nightly scans. Displacement
of individual scans by $\pm 10$~\% of a scan length results
in incorrect extrapolation of the offset scan system in edge areas and 
causes the problems discussed in Sect.~5.

\subsection{Magnitude-dependent errors}
For investigation of magnitude-dependent systematic errors
in $x$, $y$ and $v$ data we carried out a preliminary processing of 
equivalent scans with  the Tycho2 catalogue.
Direct inspection of the KMAC1-Tycho2 residuals clearly indicated
the presence of errors dependent on  magnitude.
Systematic effects in positions were found to be within $\pm 0.03''$
for the 9.5--13 magnitude range, sharply increasing to   $\pm 0.2''$ at 
$v<9$ mag. A much larger trend is seen in the KMAC1
photometry for bright $v<9.5$~mag stars where it exceeded 1.0~mag.
Systematic components of KMAC1-Tycho2 residuals were treated as 
errors in the MAC data, therefore, 
$x$, $y$ and $v$ values were corrected by subtracting  systematic 
trends.  Calibration was applied only to stars 
in the Tycho2 magnitude range V$<13$~mag.

It was considered that initial estimates of KMAC1-Tycho2 residuals 
are  somewhat
biased due to redistribution of magnitude-dependent errors 
between  
reference stars in the field. Therefore, to extract better estimates
of the magnitude-related errors from the KMAC1-Tycho2 residuals,
calculations were refined in an iterative manner.

After calibration, the residual trend in corrected positions,  
estimated using the CMC13, does not exceed $\pm 0.04''$ for the entire
magnitude range.

%%%%%%%%%%%%%%%%%%%%%%%%%%%%%%%%%%%%%%%%
\subsection{Seasonal variations of magnitude-related errors}
\begin{figure}[tbh]
\centerline{%                       
\begin{tabular}{@{}c@{}}
%\resizebox{\hsize}{!}{\includegraphics[height=120pt]{r4a.ps}} \\
%\resizebox{\hsize}{!}{\includegraphics[height=120pt]{r4b.ps}}
\includegraphics*[height=120pt, width=8.0cm]{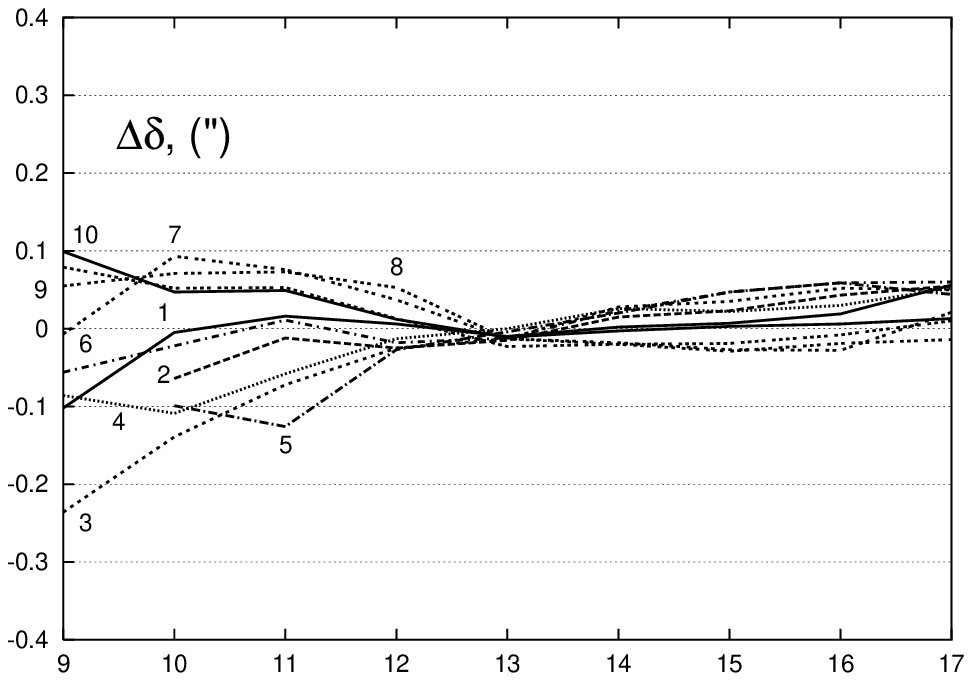} \\
\includegraphics*[height=120pt, width=8.0cm]{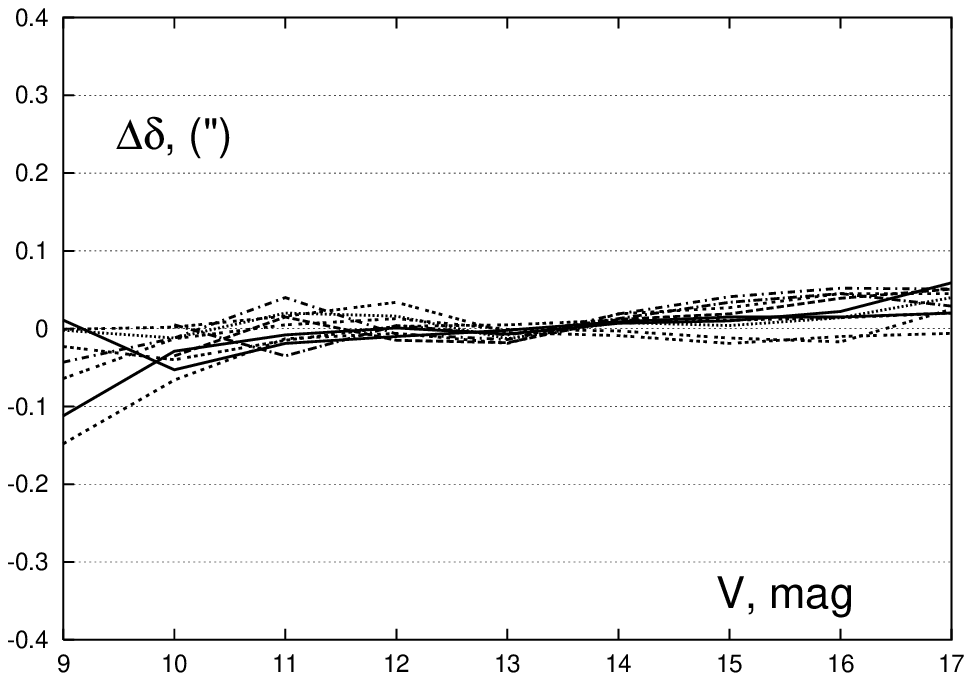} 
\end{tabular}}
\caption{ Systematic
differences KMAC1-CMC13 in DEC as a function of magnitude
for 10 groups of star fields ordered by RA; 
before (upper panel) and after (bottom) calibration (\ref{eq:r4})}
\label{ses}
\end{figure}

More explicit analysis revealed a variation  
of the magnitude error in declination with season. 
For the study we used $\Delta \delta$
differences KMAC1-CMC13 obtained with a preliminary data processing based on 
the CMC13 as a reference. Fig.~\ref{ses} shows the systematic
trend of the differences $\Delta \delta$ for 10 groups of star fields with a 
numbering that 
corresponds to their arrangement in RA, or  season.
Variations with  season are especially strong at bright magnitudes;
the difference between "winter" (numbers 1--6) and "summer" fields (7--10)
is about $0.2''$ for $v<12$~mag stars and is small at $v \approx 13$~mag.
Note that an attempt  to resolve this seasonal effect,
referring the MAC data directly
to Tycho2 and then using the KMAC1-Tycho2 differences, lead
to inconclusive results   due to the narrow magnitude range, insufficient
statistics, and, especially, the filtering effect produced
by the reduction procedure.

Considering that the picture shown in Fig.~\ref{ses} may originate from errors
in the CMC13, we used this information 
indirectly, to assume a possibility of a specific error
in the MAC declinations (or in $x$ values), and to define a function 
\begin{equation}
\label{eq:r4}
\Delta_{x}=\left \{  \begin{array}{ll}
(\beta \mbox{sin}\alpha + \gamma \mbox{cos}\alpha)(13-v) &, v<13 \\
0                                                        &, v>13 \\
\end{array} \right.
\end{equation}
that models the bias in $x$. Model 
parameters $\beta $ and $\gamma$ were found as those whose use
yielded the least square values of KMAC1-Tycho2 differences in the DEC.
The function (\ref{eq:r4}) is defined only for bright
$v<13$~mag stars; the positions of fainter stars cannot be corrected. 
Calculations yielded a solution $\beta=0.045''/$mag,
$\gamma=-0.031''/$mag with an uncertainty of $\pm 0.010''/$mag 
in each parameter.

The effect of  calibrations based on the Tycho2 catalogue is 
seen in  Fig.~\ref{ses} 
where the KMAC1-CMC13 differences after correction
are shown in the bottom panel; the residual
variations of the magnitude-related errors in DEC
are shown to be reduced to  $\pm 0.03''$ or less 
at $v \geq 10$ mag.

%% file: 2573_4.tex
\section{ Calibration of the magnitude scale}
Star magnitudes V of the KMAC1 have been computed using measured
$v$ values  corrected for instrumental and 
magnitude-dependent errors as described above.
The zero point of the V magnitude scale was
determined 
using the Tycho2 photometry of bright V$<13$~mag stars. 
The problem consisted of verification of the  magnitude 
scale linearity at its faint end,
which  cannot be directly controlled due to the
absence of faint all-sky standards in the V band.
The study and the following calibration used indirect methods 
relied upon red $r'$
and infrared J data taken from the CMC13 and 2MASS global catalogues
respectively.
The UCAC2 catalogue, as an alternative $r'$-like data source, was not
utilized since its magnitudes are only approximate and not
calibrated. Our attempt to take advantage of this
catalogue photometry resulted in a similar but slightly 
less accurate calibration
compared to that  provided using  the CMC13.

The development of the calibration model and its velidation was 
based on a photometric study of the open cluster NGC 2264 scans
obtained at the MAC, specially for this purpose.

\subsection{The open cluster NGC 2264}
For this study we compared KMAC1 V values  computed for stars of
the open cluster NGC 2264 with those given in several 
high-accuracy photometric
catalogues. As photometric V$_{st}$ standards, we used data provided by
Sung et al. (\cite{sung}) for 329 identified stars; 
Kuznetsov et al. (\cite{Kuzn}) identified 40 stars ; and 
the WEBDA Internet database of UBV CCD observations in 
open clusters provided 523 stars (Mermilliod \cite{webda}). 
First, we verified
that there are no systematic dependences of 
$\Delta$V=V--V$_{st}$ 
residuals either on B-V color or the star CCD $x$ position.

\begin{figure}[htb]
\resizebox{\hsize}{!}{\includegraphics{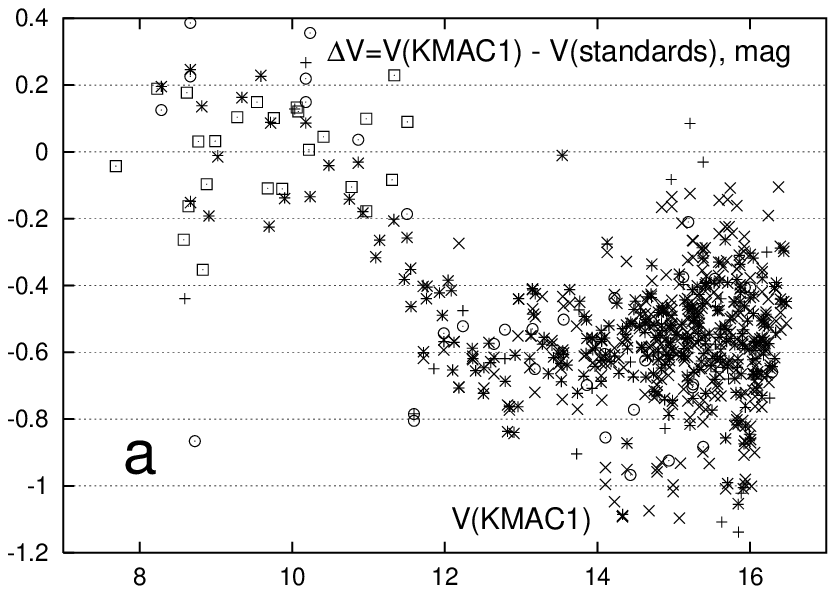}}
\resizebox{\hsize}{!}{\includegraphics{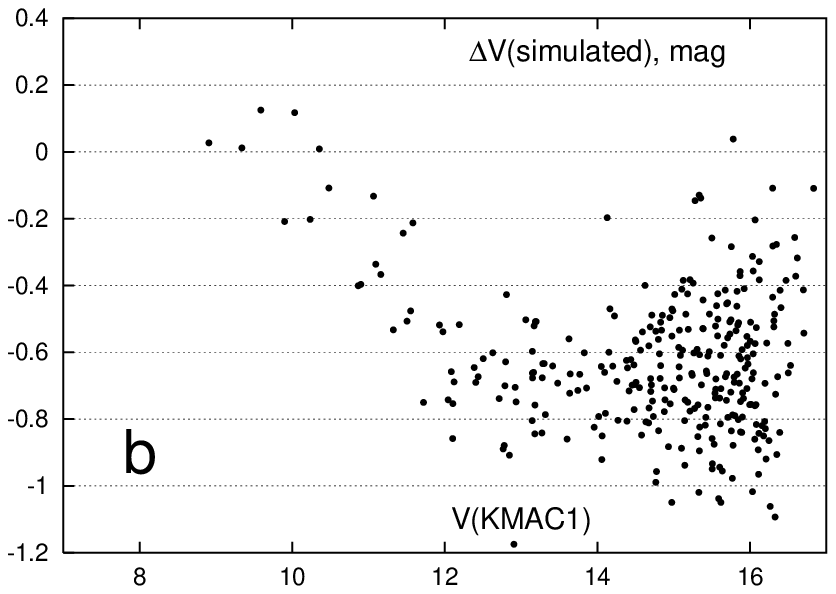}}
%\centerline{\includegraphics*[width=7.5cm]{fot1.eps}}
%\centerline{\includegraphics*[width=7.5cm]{fot2.eps}}
\caption{Systematic effect in measured V values:
{\bf a} --  found from a comparison to photometric data by
Sung et al. (\cite{sung}) (crosses), Kuznetsov et al. (\cite{Kuzn})
(circles), Internet WEBDA database (inclined crosses) and Tycho2 (squares);
{\bf b} -- the bias simulated with the model (\ref{eq:m2})
}
\label{m1}
\end{figure}
%%%%%%%%%%%%%%%%%%%%%%%%%%%%%%
Examination of $\Delta$V residuals, however, revealed a large 0.6~mag 
systematic bias of measured data for faint V$>12$~mag stars, 
shown in Fig.~\ref{m1}a. Interpretation of this plot should take
into account that
images of stars brighter than 12~mag are oversaturated
and their fluxes determined by the centroiding 
procedure can be systematically biased.
The resulting  effect in magnitude is however opposite 
because the zero point of the V scale is referred to bright Tycho2 stars.
Another important feature of 
the plot is the linearity of magnitude scales in
either bright V$<10.5$~mag and faint V$>12$~mag segments of the V-axis,
however, with different zero points.

\begin{figure}[htb]
%\resizebox{\hsize}{!}{\includegraphics{fotn1.eps}}
%\resizebox{\hsize}{!}{\includegraphics{fotn2.eps}}
\centerline{\includegraphics*[width=8.4cm]{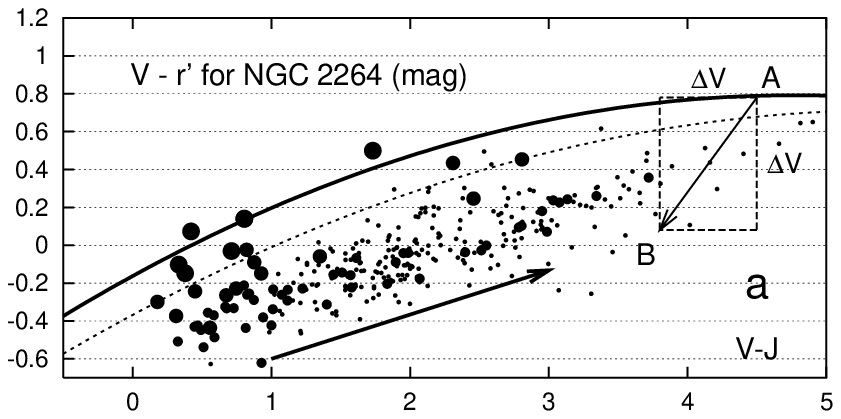}}
\centerline{\includegraphics*[width=8.4cm]{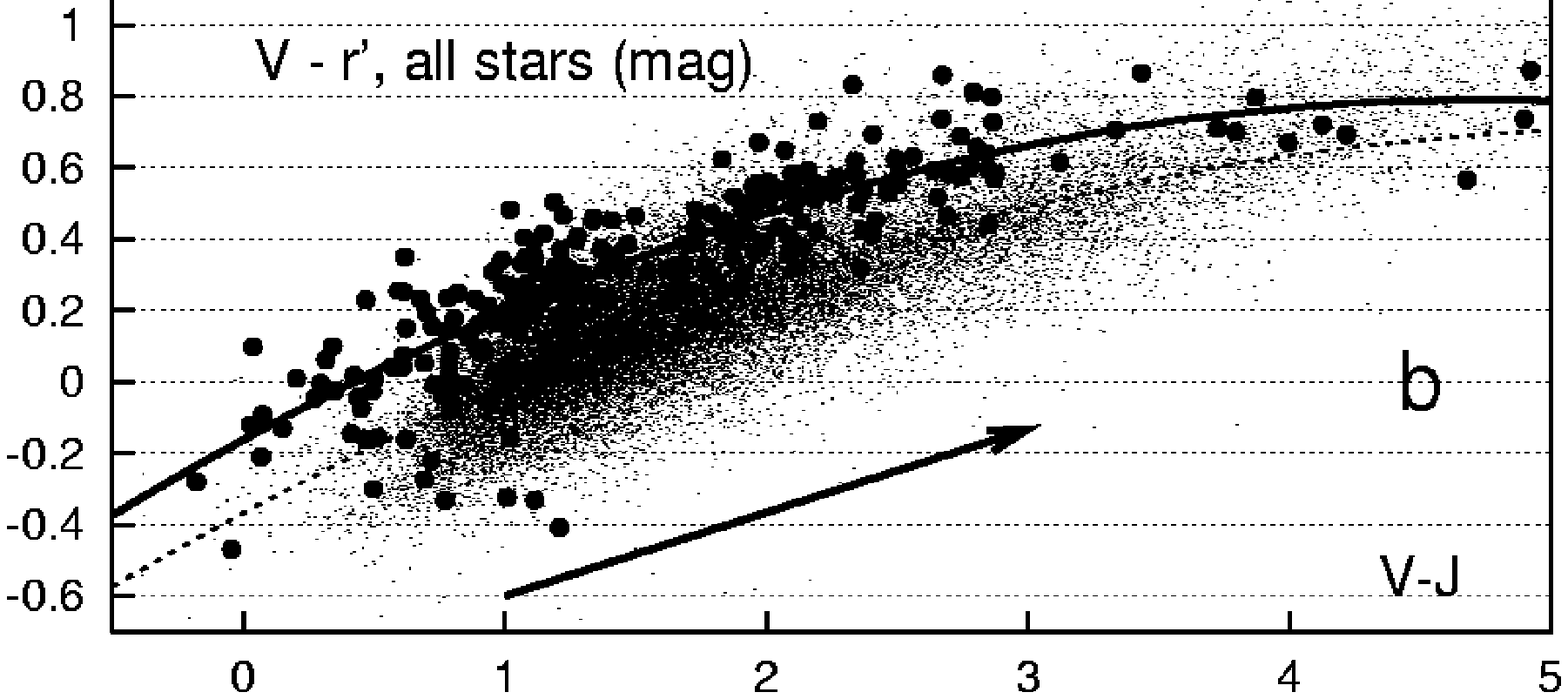}}
\caption{Two-color diagrams V$-r' \sim$V--J:
{\bf a} -- for NGC 2264 stars, symbol size indicates star brightness 
(9 to 16~mag);
{\bf b} -- for all stars, large dots show V$<$10.5~mag stars.
Solid line (\ref{eq:m1}) fits the bright star location, the dashed line
refers to all stars, thick arrows show the direction of interstellar reddening.
A geometry explaining the model (\ref{eq:m2})
is also shown (see text)
}
\label{m2}
\end{figure}

We tried to simulate this systematic effect using two-color 
V$-r' \sim$V--J diagrams that were built for NGC 2264 (Fig.~\ref{m2}a) and for
a complete list of the KMAC1 stars 
(Fig.~\ref{m2}b). Star distributions in both plots are clearly 
separated depending on V, bright stars being shifted systematically 
upward relative to faint stars. The shift  
does not depend on  V--J color, so we refer it entirely to
magnitude-dependent errors in  MAC photometry. 
A number of other two-color
diagrams were also tested, including those that incorporate H and K 
infrared data
from the 2MASS catalogue;  it is the V$-r' \sim$V--J
diagram that ensures the best separation of stars in  V.

%%%%%%%%%%%%%%%%%%%%%%%%%%%%%%
The differences V$-r'$ for bright KMAC1 stars  are approximated
by a function (solid line in Fig.~\ref{m2}b)
\begin{equation}
\label{eq:m1}
f_{V-J}= \overline{V-r'}=-0.163+0.4016(V-J)-0.0422(V-J)^2
\end{equation}
A small residual scatter of $\pm 0.114$~mag suggests that the 
distribution is uniform and the dependency
(\ref{eq:m1}) is valid for any  field. In particular, 
the function
$f_{V-J}$ matches reasonably well the bright star distribution for NGC 2264 
(solid line in Fig.~\ref{m2}a). Note that the function
(\ref{eq:m1}) represents a zero-point of the V scale since 
the bright stars, for the most part,  are the Tycho2 stars used
as a reference for photometry. The dashed line refers to all stars.
The position of faint V$>12$~mag stars
are shifted downwards and the fitted dashed line
is almost parallel to $f_{V-J}$ in the most populated 0.5--2.0 area
of V--J colors. This leads to the very important conclusion that
errors $\Delta V$ in MAC photometry do not depend on the  color,
which is consistent with  previous results based on the use 
of photometric standards; rather, they are a function of V.

The calibration of  faint star photometry to the instrumental
magnitude system defined by bright stars is based on the use of the
fitting curve (\ref{eq:m1}) as reference.
Consider the two-color diagram where the
unbiased star location A is a point V$_{st}-r'$,V$_{st}-$J in the
fitting curve (Fig.~\ref{m2}a). 
The star's measured
position, B, is shifted by $\Delta V$ in both directions, to 
V$_{st}+\Delta V-r'$,V$_{st}+\Delta V-$J. This geometry allows us to
express the point B distance to the fitting curve 
(\ref{eq:m1}) in two ways, as V$-r'-\overline{V-r'}$ and as
$ \Delta V(1-f'_{V-J})$ where
$f'_{V-J}$ is the derivative. Hence we derive an estimate
\begin{equation}
\label{eq:m2}
\Delta V= V-r'-\overline{V-r'}/(1-f'_{V-J})
\end{equation}
of the  bias. In the first approximation (when $f' \approx 0$),
it is equal to the vertical distance between the measured
star location B in the diagram and the fitting color curve $\overline{V-r'}$.

Using Eq.\ (\ref{eq:m2}), we computed errors $\Delta V$ for 
 each NGC 2264 star with $r'$ and J data available. The results
shown in Fig.~\ref{m1}b  match well
the systematic trend found directly on photometric standards
(Fig.~\ref{m1}a), except for a
small systematic discrepancy of about 0.1~mag at the faint V end.

\subsection{Galactic extinction}
The above analysis does not take into account interstellar extinction, which
requires  special considerations. With respect to bright stars we may
however assume that for the most part they are  nearby objects
not affected by extinction. The distribution of bright stars in 
Fig.~\ref{m2}a therefore is expected to follow a natural temperature
reddening, at least for V-J$<3$ mag. Considering that the interstellar
reddening is inversely proportional to wavelength 
(Whitford \cite{whitford}),
we find a direction along which faint stars may move
in the color diagrams (arrows in Fig.-s~\ref{m2}a and b). The inclination
of the reddening line 0.234 was computed using the median wavelengths 
0.54, 0.623 and 1.25~nm of V, $r'$ and J filters respectively, with no
allowence for spectral class. It is seen that the temperature and
Galactic reddening are indistinguishable since the corresponding curves are
almost parallel. Thus, both the function (\ref{eq:m1}) and the reddening 
curve have equal inclination at V-J=2.0. The differential shift of faint
stars off the temperature line is therefore  small and for most stars with
V-J ranging from +1 to +2 is, on average, less than 0.05~mag,
supposing that the extinction in V does not exceed 1.5~mag (1.0~mag
color excess in V-J). Since further computations are performed 
by  averaging over the total star sample, 
the expected error in the photometric calibration is even smaller.

\subsection{Calibration of individual fields}
\begin{figure}[htb]
%\resizebox{\hsize}{!}{\includegraphics{fot3.eps}}
\centerline{\includegraphics*[width=8.0cm]{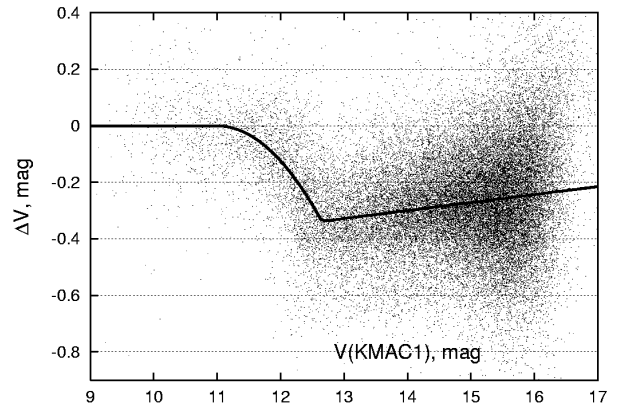}}
\caption{Individual estimates (\ref{eq:m2}) of   $\Delta V$ bias 
in  magnitudes for all stars, and its approximation (\ref{eq:m3})
by the solid line}
\label{m3}
\end{figure}

The point estimates of Eq.\ (\ref{eq:m2}) thus form the basis for 
the calibration of the magnitude scale in isolated fields
with the use of external color information from  CMC13 and 2MASS.
Taking into account the insufficient statistics
and the complicated shape of the bias to be removed,
we introduced for  each field  a simple one-parameter
model that represents the systematic dependence of $\Delta V$ 
as a function of V. The form of this function $\overline{\Delta V(V)}$ was
chosen considering the distribution of $\Delta V$ values computed with 
Eq.\ (\ref{eq:m2}) for all stars with known $r'$ and J values (Fig.~\ref{m3}).
The overall distribution of  $\Delta V$
is like that shown in Fig.~\ref{m1}a for  NGC 2264, except with more
shallow knee, and is fitted with the function 
\begin{equation}
\label{eq:m3}
\overline{\Delta V(V)}= \left \{   \begin{array}{ll}
		0, & V<11\\
		\nu [(V-11)/(1.65)]^2, & 11<V<12.65 \\
                \nu +0.029(V-12.65), & V>12.65
\end{array}
		\right.
\end{equation}
where the single $\nu$ parameter is the bias magnitude at
V=12.65~mag.
The fit obtained with $\nu=-0.340$ is shown by the solid line.
%%%%%%%%%%%%%%%%%%%%%%%%%%%%%%
\begin{figure}[htb]
%\resizebox{\hsize}{!}{\includegraphics{2573fg13.eps}}
\centerline{\includegraphics*[width=8.0cm]{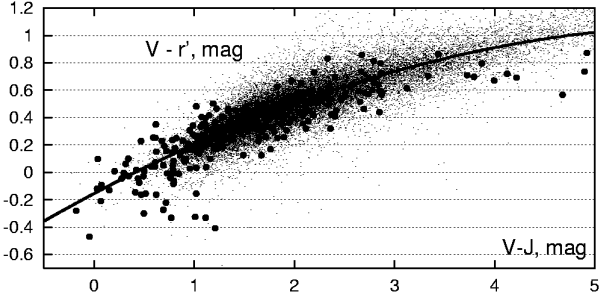}}
\caption{V$-r' \sim$V--J color diagram  for all stars, 
with V corrected. The solid line is the fitting function (\ref{eq:m4})
}
\label{m4}
\end{figure}
%%%%%%%%%%%%%%%%%%%%%%%%%%%%%%

Further analysis has shown that the $\nu$ value is to be determined for
each star field individually, by fitting estimates (\ref{eq:m2}) with
the model (\ref{eq:m3}). The calibration of  magnitudes therefore was 
performed with individual $\nu$ values that varied from
-0.63 to -0.14 (-0.58 for NGC 2264). The scatter of $\nu$ values,
in particular, is the cause of the large $\pm 0.157$~mag dispersion of points
in Fig.~\ref{m3}.

\begin{figure}[htb]
%\resizebox{\hsize}{!}{\includegraphics{fot5.eps}}
%\resizebox{\hsize}{!}{\includegraphics{cfbordo.eps}}
\centerline{\includegraphics*[width=8.0cm]{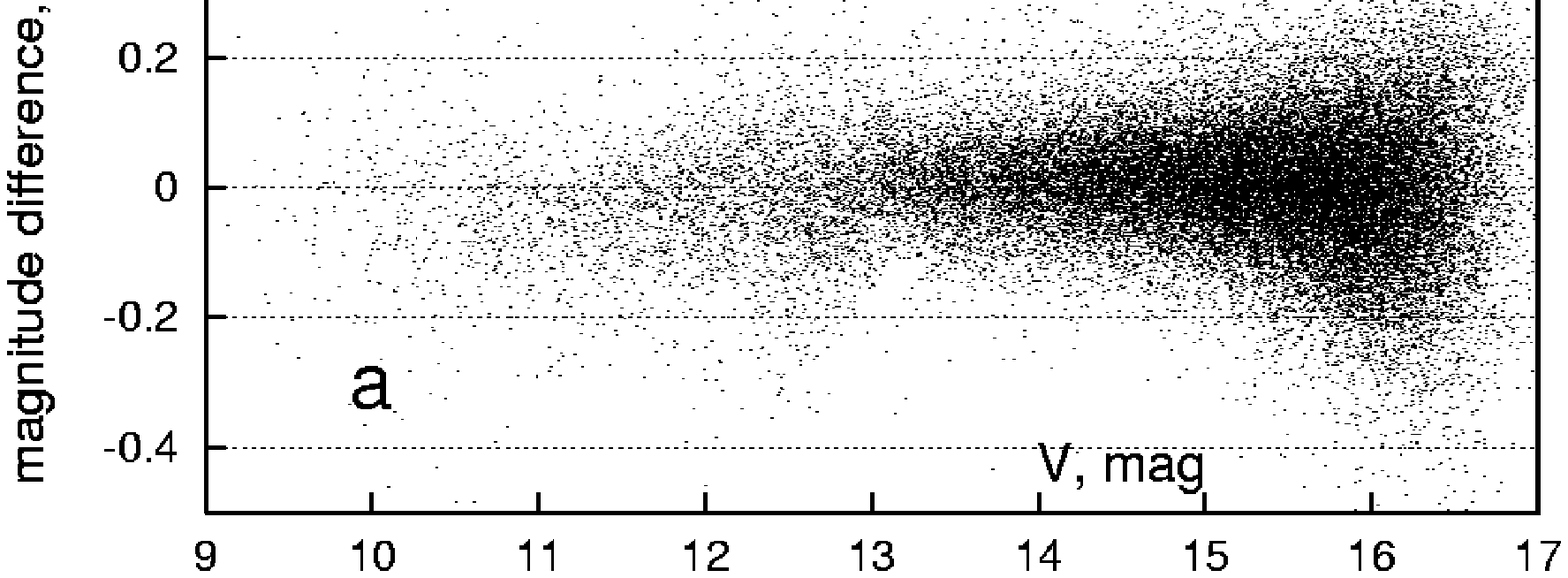}}
\centerline{\includegraphics*[width=8.0cm]{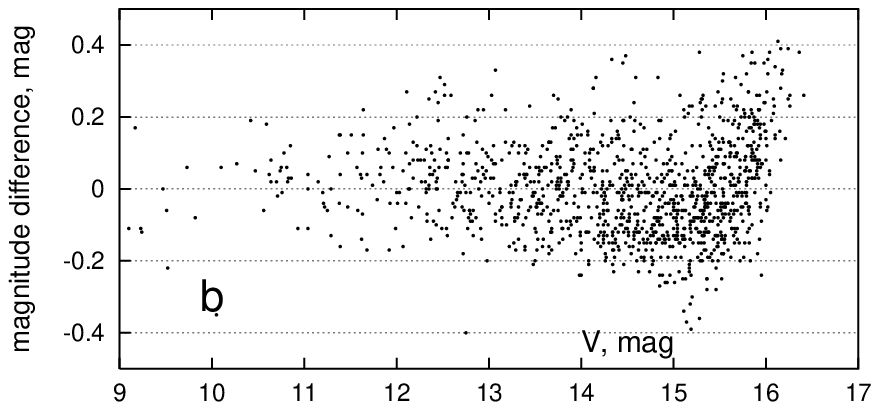}}
\caption{Magnitude residuals as a function of V: 
{\bf a} --  V$-r'$ differences 
measured with reference to the fitting curve shown in Fig.~\ref{m4};
{\bf b} --  residuals between the KMAC1 and the Valinhos catalogue V values. 
           }
\label{m5}
\end{figure}

The efficiency of corrections is seen from
the two-color V$-r' \sim$V--J  diagram built
with final calibrated V values (Fig.~\ref{m4}). The relative shift
of bright to faint stars like that shown in Fig.~\ref{m2}b
was eliminated; also, the standard deviation of points from the fitting
curve
\begin{equation}
\label{eq:m4}
V-r'=-0.147+0.381(V-J)-0.0293(V-J)^2
\end{equation}
was improved from $\pm 0.114$~mag to $\pm 0.087$~mag. 
The scatter of  V$-r'$ residuals from the fitting curve (\ref{eq:m4}),
plotted in Fig.~\ref{m5}a as a function of V, 
includes errors in $r'$ values and indicates the upper limit
of KMAC1 magnitude errors. This plot also shows a good
elimination of systematic errors.

Fig.~\ref{m5}b shows a comparison of the KMAC1 and Valinhos meridian 
circle photometry (Camargo et al. \cite{camargo}) for 1190 
stars in 13 stellar fields. 
The residuals contain no large systematic trend; the
standard deviation of data points is $\pm 0.13$~mag.
Consideration of local fields, however, indicates
local systematic discrepancies in magnitudes 
sometimes reaching $\pm 0.10$~mag, with a random scatter of residuals  
of about $\pm 0.10$~mag. 
Considering  the accuracy of the Valinhos photometry, which is
about $0.10$~mag (Viateau et al. \cite{vit}), 
we estimate that the KMAC1 data
is of the same or better accuracy.

%%%%%%%%%%%%%%%%%%%%%%%%%%%%%%

In conclusion, we give a relation between V and the CMC13 $r'$ values
using $r'-$J  colors:
\begin{equation}
\label{eq:m5}
V-r'=-0.015+0.376(r'-J)-0.0269(r'-J)^2
\end{equation}

%% file: 2573_5.tex
\section{ Reduction to the ICRF}
\begin{figure}[htb]
%\centerline{\includegraphics*[bb = 85 55 683 440, width=10.0cm]{fg5.eps}}
\begin{tabular}{@{}c@{}c@{}}
\includegraphics*[width=4.2cm]{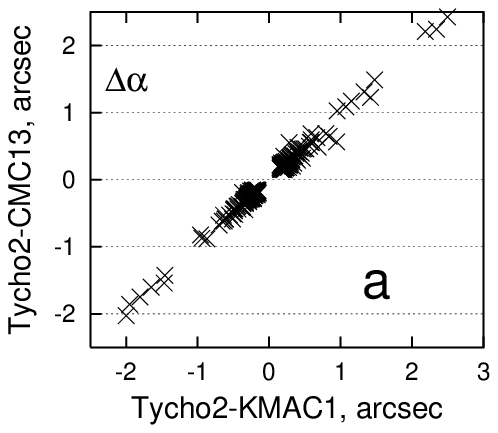} & \includegraphics*[width=4.2cm]{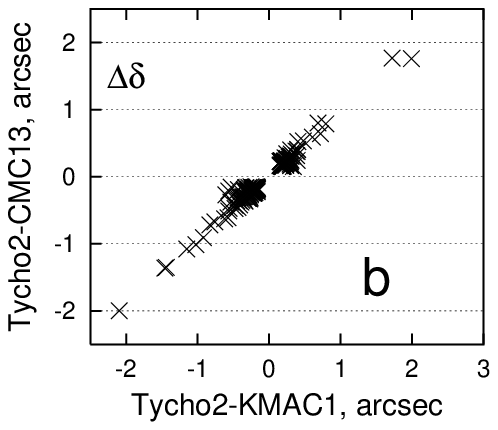}\\
\includegraphics*[width=4.2cm]{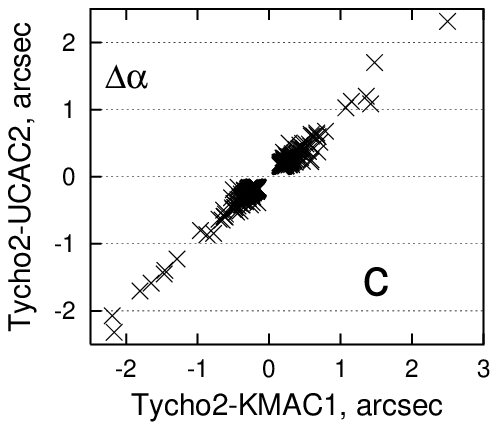} & \includegraphics*[width=4.2cm]{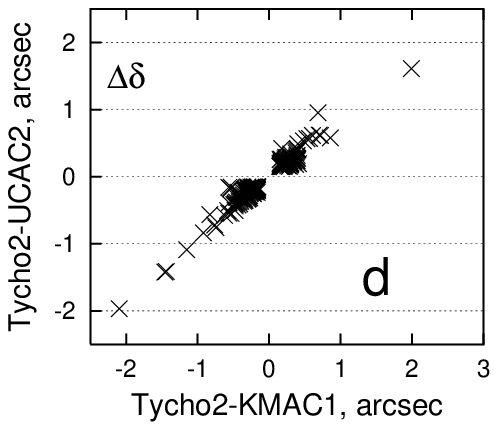}\\
\end{tabular}
\caption{Correlation between Tycho2-CMC13 
 and  Tycho2-KMAC1  preliminary differences:
{\bf a} - in RA; {\bf b} - in DEC; 
correlation between Tycho2-UCAC2  and  Tycho2-KMAC1  preliminary differences:
{\bf c} - in RA; {\bf d} - in DEC; only differences exceeding 0.15$''$ 
are shown}                        
\label{tycho}
\end{figure}

Compilation of KMAC1 started from astrometric calibration of the
measured data as described in the previous Sections. Conversion of corrected 
CCD  $x$, $y$ positions to equatorial coordinates originally was intended
to be performed  using the Tycho2 catalogue which is the best
optical representation of the ICRF. Prior to this conversion
we performed a tentative study to determine how well 
the positions of the Tycho2 catalogue match the 
modern catalogues CMC13, UCAC2 
and our observations. 
For this purpose we selected Tycho2 stars that:
   \begin{itemize}
      \item are located in the fields observed;
      \item are listed at least in either the CMC13 or UCAC2;
      \item  have preliminary KMAC1 positions that
	do not differ from those in either the CMC13 or UCAC2 by
more than $0.4''$;
      \item have proper motions $\mu_{\alpha}$ and $\mu_{\delta}$  not
exceeding $\pm 0.3''$/year.
   \end{itemize}         
Fig.~\ref{tycho} presents individual  
Tycho2-CMC13  and Tycho2-UCAC2  differences plotted versus Tycho2-KMAC1 
preliminary differences for 1843 Tycho2 stars that meet the above 
conditions. The almost diagonal location of data points testifies to the
good agreement between  the ground-based catalogues,
which provide positions normally consistent to $0.1$---$0.2''$. Very large
deviations, $>\pm 1''$ , usually seen for faint  stars V$>$12~mag,
seem to
originate from errors in the Tycho2 positions. 
  For 1.4 percent
of stars, positional errors of  Tycho2 exceed $\pm 0.5''$, and in about 12\% 
they are larger than $\pm 0.2''$.  The r.m.s. of the difference
of "Tycho2 - CCD catalogue" in RA is 204, 166 and 210~mas, 
respectively, for the CMC13, UCAC2 and KMAC1. In DEC these estimates are
148, 123 and 170~mas. Somewhat  minor  deviations,
149~mas in RA and 129~mas in DEC, were found by
Camargo et al. (\cite{camargo})
from the analysis of the Valinhos transit circle observations.
Of course, the values cited include comparison catalogue errors which are
usually not large.

The above analysis establishes the 
degradation of the Tycho2 data  at the epoch of  KMAC1 observations, 
probably  due to uncertanties
in proper motions. This problem is often allowed for in different ways
when referring CCD observations to  equatorial coordinates.
Thus, at the Flagstaff Astrometric Scanning Telescope  reductions are
made by applying weights to the Tycho2 stars depending on their brightness
(Stone et al. \cite{fasst}).

%%%%%%%%%%%%%%%%%%%%
%%%%%%%%%%%%%%%%%%%%%%%%%%%%%%%%%%%%%%%%%%

This discussion suggests
that a reliable reduction to the ICRF using the Tycho2 catalogue requires
the use of a sufficiently large number of reference stars which are to be
first filtered or weighted to eliminate problem stars.
This is especially important in our case since, due to the rather short 
scan length,
some fields in sky areas with a low star density are represented only
by 6--8 Tycho2 stars. 
Also, the accuracy of the reduction is  affected by inhomogeneity in
the sky distribution of reference stars whose images,
in addition, are oversaturated and poorly measured. 
At the first stage of referencing we therefore
detected and removed all 
Tycho2 problem stars whose positions deviated from
the CMC13, UCAC2 and preliminary KMAC1 data by more than $\pm 0.2''$.
This greatly improved the reliability of the conversion to equatorial
coordinates and was found to be more efficient than the usual 
search for outliers based on an iterative approach to the least-squares 
solution. With a truncation limit of $\pm 0.2''$, reliable results
were obtained, however, for 106 fields only. 
For another 53 fields, 
a good transformation to the ICRF required a further rejection of 
reference stars with Tycho2-CMC13 and Tycho2-UCAC2 differences in 
the range from $\pm 0.2''$ to $\pm 0.15''$. 
For the 33 remaining fields with a low reference star density, the 
reduction was found to give
quite unstable and ambiguous solutions highly sensitive to 
any changes in the reference star set.

A further comparison with the CMC13 and UCAC2 positions has shown 
that large systematic deviations are present at the edges of some fields.
This  concerns those fields that at some nights were observed
with incorrect telescope pointings (made by
hand since the MAC is not automatic); in a few  cases
the relative displacement of sky strips exceeds $10'$ in RA.
Individual scans thus were not
exactly overlapped as was assumed at the phase of equivalent scan
formation. To eliminate this fault, the offset  regions were
truncated.
                             
A rigorous conversion to the ICRF using the Tycho2
catalogue was achieved for 159 sky fields, most of which had
a high star density. 
Conversion for the complete data array 
(192 sky fields) required 
the use of the CMC13 and UCAC2 catalogues which are known to be in 
the  ICRF system. The reduction was performed with well-measured stars 
not fainter than 14.5~mag and by limiting their number to 170. 
Reference catalogues were used in a combined form, 
with equal weights. No truncation 
of offset scan edges was applied since the large number of
reference stars ensured very tight referencing using
spline fitting.

Thus, the catalogue KMAC1  exists in the two
versions:  with reduction to the Tycho2 (KMAC1-T)
and  to the CMC13 and UCAC2 catalogues (KMAC1-CU).
No rejection of stars with large deviations from comparison catalogues
was applied. V magnitudes given in both catalogues are identical and
based on the Tycho2 photometry with the corrections described in Sect.~4.

\section{Proper motions}
The first epoch positional data used for the computation of proper motions
was taken from the USNO-A2.0 catalogue. However the epoch difference of about
50 years prevented direct identification of stars with large
proper motions. To improve the reliability of identification, 
the USNO-B1.0 catalogue proper
motions were used and were applied to reduce displacement of star positions
due to the difference in epochs. For stars not found
in the  USNO-B1.0, the identification was performed with no proper
motion information applied. In this case, a window
used for identification
was set from $1.4''$ to $2.4''$ depending on the star magnitude.
Stars with proper motions larger than about 40~mas/year therefore
cannot be found in the  USNO-A2.0 with no proper motion information
from the USNO-B1.0.

The percentage of KMAC1 stars supplemented with proper motions
varies from 53 to 97, and on average is 90\%. The highest ratio
of 93\% is obtained in the magnitude range from 13 to 17~mag,
dropping to 47\% for V$<$12~mag stars and to 74\% for V$>$17~mag.

Considering  the approximate precision of $\pm 250$~mas,
the mean epoch 1954 of the USNO-A2.0 and the internal positional
precision of the KMAC1 (Table~\ref{char}), we find a formal estimate
of proper motion errors of 5--6~mas/year.

%% file: 2573_6.tex
\section{Characteristics of the catalogue}
\subsection{Description of the catalogue}
\begin{figure}[tbh]
\includegraphics*[bb =  33 546 545 811, width=8.0cm]{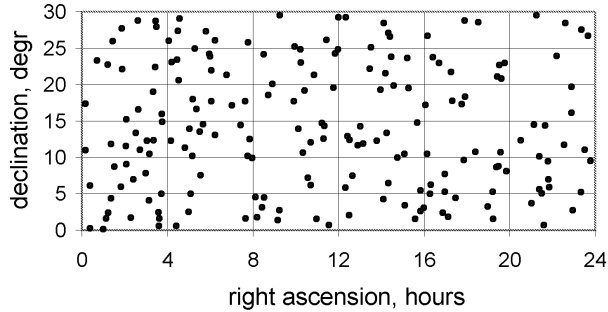} 
\caption{Distribution of KMAC1-CU fields across the sky}
\label{sky}
\end{figure}

The catalogue KMAC1, as explained above, was released in the two versions
and can be obtained in electronic form from the  CDS or via 
anonymous ftp://ftp.mao.kiev.ua/pub/users/astro/kmac1.
The KMAC1-CU catalogue contains 115\,032 stars in 192 sky fields and is
referred to the  CMC13 and UCAC2; the KMAC1-T contains 104\,796
stars (91\% of the total star number) 
in 159 fields and is referred to the Tycho2 catalogue. 
The location of  KMAC-CU fields in the 
sky is shown in Fig.~\ref{sky}. All fields
are located in a declination zone from 0$^{\circ}$ to 30$^{\circ}$; 
the mean epoch of observations
is  2002.33. The main
characteristics of the catalogue are given in Table~\ref{char}.

%%%%%%%%%%%%%%%%%%%%%%%%%%

\begin{table}[tbh]
{\footnotesize
\caption{\small Main characteristics of catalogues
KMAC1-T and KMAC1-CU}
\label{char}
\begin{center}
\begin{tabular}{@{}l@{}ll@{}}
\hline
Catalogue version	&      KMAC1-T  	 &	  KMAC1-CU\rule{0pt}{11pt} \\
Reference catalogues     &        Tycho2          &        CMC13, UCAC2 \\
Number of fields	&    	  159            &         192     \\
Declination zone  & 0$^{\circ}$ to +30$^{\circ}$ & 0$^{\circ}$ to +30$^{\circ}$ \\
Number of stars         &          104796            &       115032    \\
Precision of positions   &                     &                      \\
V$<14$~mag:              &     30--50~mas$^*$  &     30--50 mas$^*$   \\
                         &     50--80 mas$^{**}$ &  30--70 mas$^{**}$  \\               
V=16~mag:                &    170~mas$^*$     &    170 mas$^*$\\
                         &      180 mas$^{**}$  &    170 mas$^{**}$  \\
Formal precision of      &                      &                      \\
proper motions:         &      5--6~mas/year   &       5--6~mas/year     \\
Precision of photometry &                         &                 \\
V$<15$~mag:             &     0.02--0.04~mag$^{*}$ & 0.02--0.04~mag$^{*}$  \\
                       &     0.06--0.08~mag$^{**}$ & 0.06--0.08~mag$^{**}$  \\
\hline
\end{tabular}
\end{center}
\begin{tabular}{l}
$^*$)  internal errors; \, \,   \, \, $^{**}$)  external errors  \\
\end{tabular}
\begin{center}
\begin{tabular}{l}
Astronomical data included: 
$\alpha$, $\delta$, $\mu_{\alpha}$ , $\mu_{\delta}$, 
B, V, R, r$'$, J \\
\hline
\end{tabular}
\end{center}
}
\end{table}

Besides the original positions given at the epoch of observations, 
proper motions and original V values, the
KMAC1 catalogue  contains B, R values from the USNO-B1.0 
for 83\% of the stars identified; r$'$ values are taken
from the CMC13 for 67\% of stars and J values from the 2MASS catalogue
available for 94\% of stars.
Usual supplementary information, including internal error estimates,
the number and epoch of observations, the image quality index, image size for
extended objects and cross-identification to
the USNO-B1.0 is also given.

Note that the flagged image quality index (see Sect.~2)
may indicate centroiding problems of various origins (e.g. binary or
unresolved stars);
the catalogue positions therefore are probably biased. The unequal number of
observations for RA and DEC data means that a rejection of bad
measurements was applied which also may indicate certain problems with
image quality not marked in the image quality index. 
These stars should not be used when very high accuracy of positions is required.

\begin{figure}[tbh]                                                                                                                                      
\includegraphics*[width=8.0cm]{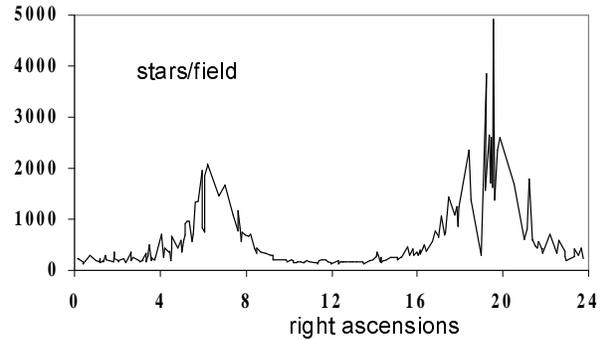}                                                                                                               
\caption{Distribution of KMAC1 star number per field on right ascensions}                                                                                                      
\label{Gal}                                                                                                                                              
\end{figure}

The list of sky strips, the IAU designations of the central ICRF object
and star numbers $N_T$ and $N_{CU}$
containing, respectively, in the KMAC1-T and KMAC1-CU for each strip,
are given in Table~\ref{destbl}. 
Note that  $N_T$ is often lower than $N_{CU}$ due to a truncation of sky strip
edges applied to some  fields (Sect.~5).
The star number distribution over
fields is highly inhomogeneous and depends on the Galactic latitude.
This distribution as a function of RA is shown in 
Fig.~\ref{Gal}.

\begin{table*}[htb]
\caption{\ Fields with ICRF objects and star numbers contained 
in the KMAC1-CU and KMAC1-T catalogues}
{\scriptsize  %\tiny %\scriptsize
\label{destbl}
\begin{tabular}{rrr|rrr|rrr|rrr}
\hline      
Identifier   &$N_{CU}$ &$N_T$&Identifier   &$N_{CU}$ &$N_T$&Identifier   &$N_{CU}$ &$N_T$&Identifier   &$N_{CU}$ &$N_T$\rule{0pt}{11pt} \\
\hline                                                                            
001031.0+105829 & 274 &258   &044907.6+112128 & 462 & 462  & 105829.6+013358 & 179 & --   &164125.2+225704 & 591 &  591\rule{0pt}{11pt}\\ 
001033.9+172418 & 243 &232   &045952.0+022931 & 602 & 602  & 111358.6+144226 & 164 & --   &165259.3+022358 & 798 &  798  \\ 
002232.4+060804 & 190 &190   &050145.2+135607 & 366 &  366 & 111857.3+123441 & 162 &162   &165809.0+074127 & 701 &  701  \\ 
002225.4+001456 & 127 & --   &050523.1+045942 & 621 &  602 & 112027.8+142054 & 139 & --   &165833.4+051516 & 685 &  675  \\ 
004204.5+232001 & 308 &308   &050927.4+101144 & 715 &  715 & 112553.7+261019 & 219 & 219  &170734.4+014845 &1105 & 1105  \\ 
005905.5+000651 & 183 & --   &051002.3+180041 & 939 &  903 & 113320.0+004052 & 222 & --   &171521.2+214531 & 734 &  734  \\ 
010838.7+013500 & 167 & 167  &051601.9+245830 & 985 &  985 & 114505.0+193622 & 220 & 220  &171913.0+174506 & 776 &  763  \\ 
011205.8+224438 & 239 & 222  &052109.8+163822 & 984 &  984 & 115019.2+241753 & 193 & --   &172824.9+042704 &1483 & 1483  \\ 
011343.1+022217 & 171 & 171  &053056.4+133155 & 583 &  568 & 115825.7+245017 & 145 & --   &174535.2+172001 &1130 & 1087  \\ 
012141.5+114950 & 196 & 194  &053238.9+073243 & 741 &  741 & 115931.8+291443 & 131 & 131  &175132.8+093900 &1305 & 1290  \\ 
012156.8+042224 & 170 & 159  &053942.3+143345 &1376 & 1338 & 121923.2+054929 & 213 & 213  &175342.4+284804 & 936 &  936  \\ 
012642.7+255901 & 303 & 303  &054734.1+272156 &1402 & 1375 & 122006.8+291650 & 115 & --   &175559.7+182021 &1256 & 1153  \\ 
013027.6+084246 & 223 & 223  &055704.7+241355 &1998 & 1927 &  122503.7+125313 & 192 & 192 &182402.8+104423 &2426 & 2426  \\ 
014922.3+055553 & 188 & 182  &055932.0+235353 & 925 &  --  &  122906.6+020308 & 198 & --  &183250.1+283335 &1419 & 1419  \\ 
015127.1+274441 & 360 & 327  &060309.1+174216 & 768 &  768 &  123049.4+122328 & 180 & 180 &185802.3+031316 & 301 &  301  \\ 
015218.0+220707 & 257 & 257  &060351.5+215937 &1911 & 1776 &  123924.5+073017 & 197 & 197 &191306.8+013423 &3941 & 3941  \\ 
020346.6+113445 & 192 & 192  &061350.1+260436 &2077 & 1836 &  125438.2+114105 & 164 & 164 &191254.2+051800 &1600 & 1600  \\ 
020434.7+090349 & 176 & 176  &061357.6+130645 &2128 & 2128 &  130020.9+141718 & 184 &147  &192218.6+084157 &2755 & 2755  \\ 
020450.4+151411 & 195 & 195  &064524.0+212151 &1488 & 1488 &  130933.9+115424 & 189 &--   &192559.6+210626 &1755 & 1755  \\ 
021748.9+014449 & 247 & 159  &070001.5+170921 &1723 & 1723 &  132700.8+221050 & 180 &180  &192840.8+084848 &2682 & 2682  \\ 
022428.4+065923 & 148 & --   &072516.8+142513 &1121 & 1121 &  133037.6+250910 & 150 &--   &193124.9+224331 &1701 & 1701  \\ 
023145.8+132254 & 268 & 192  &073807.3+174218 & 782 &  782 & 134733.3+121724 & 202 &202   &193435.0+104340 &5024 & 5024 \\ 
023752.4+284808 & 372 & 350  &073918.0+013704 &1216 & 1175 & 135704.4+191907 & 227 &227   &193648.0+205136 &1405 & 1405 \\ 
023838.9+163659 & 199 & --   &074533.0+101112 & 840 &  840 & 140501.1+041535 & 280 &280   &194606.2+230004 &2414 & 2414 \\ 
024229.1+110100 & 258 & --   &074625.8+254902 & 578 &  559 & 140700.3+282714 & 365 &--    & 195005.5+080713 &2668 & 2668 \\ 
025927.0+074739 & 201 & 188  &075052.0+123104 & 759 &  756 & 141154.8+213423 & 318 &249   & 203154.9+121941 &1739 & 1700 \\                         
030230.5+121856 & 166 & --   &075706.6+095634 & 708 &  708 & 141558.8+132023 & 170 &170   & 210138.8+034131 & 608 &  608 \\ 
030826.2+040639 & 189 & 172  &080757.5+043234 & 672 &  513 & 141908.1+062834 & 295 &291   & 210841.0+143027 & 846 &   779\\ 
030903.6+102916 & 202 & 202  &081126.7+014652 & 724 &  717 & 141959.2+270625 & 204 &184   & 211529.4+293338 &1858 &  1793\\ 
031951.2+190131 & 349 & 287  &082550.3+030924 & 431 &  431 & 142440.5+263730 & 193 &193   & 212313.3+100754 & 604 &   604\\ 
032153.1+122113 & 177 & --   &083052.0+241059 & 343 &  303 & 142700.3+234800 & 193 &174   & 212344.5+053522 & 622 &   622 \\ 
032536.8+222400 & 293 & 293  &083148.8+042939 & 451 &  447 & 143439.7+195200 & 209 &196   & 213032.8+050217 & 478 &   478 \\ 
032635.3+284255 & 361 & 361  &084205.0+183540 & 364 &  282 & 144516.4+095836 & 283 &283   & 213638.5+004154 & 464 &   438 \\ 
032957.6+275615 & 525 & 494  &085448.8+200630 & 349 &  244 & 150424.9+102939 & 267 & 259  &213901.3+142335 & 607 &   607  \\ 
033409.9+022609 & 182 & 182  &090910.0+012135 & 312 &  --  & 150506.4+032630 & 239 & 239  &214710.1+092946 & 433 &   433  \\ 
033647.2+003516 & 252 & --   &091437.9+024559 & 307 &  --  & 151340.1+233835 & 271 & 254  &214805.4+065738 & 448 &   448  \\ 
033717.1+013722 & 241 & 241  &091552.4+293324 & 223 & 196  & 151656.7+193212 & 272 & 272  &215137.8+055212 & 330 &   330  \\
034328.8+045802 & 211 & 211  &095456.8+174331 & 217 &  --  & 153452.4+013104 & 496 & 496  &221205.9+235540 & 734 & 696    \\
034423.1+155943 & 219 & --   &095649.8+251516 & 177 & 177  & 154049.4+144745 & 310 & 293  &223236.4+114350 & 348 & 348    \\
034506.4+145349 & 251 & --   &100741.4+135629 & 218 & 218  &  154929.4+023701 & 448 & 448 & 223622.4+282857 & 601 & 581   \\  
040305.5+260001 & 717 & 717  &101353.4+244916 & 192 & --   &  155035.2+052710 & 302 & --  & 225307.3+194234 & 395 & 395   \\  
040922.0+121739 & 256 & --   &101447.0+230116 & 156 & 156  &  155930.9+030448 & 383 &  374& 225357.7+160853 & 278 & 241   \\  
041243.6+230505 & 473 & 418  &102010.0+104001 & 148 & 148  &  160332.0+171155 & 297 &  297&  225717.5+024317 & 202 &  --  \\  
042446.8+003606 & 375 & 375  &102444.8+191220 & 168 & 165  &  160846.2+102907 & 412 &  412&  232044.8+051349 & 290 &  266 \\  
042655.7+232739 & 397 & 388  &103334.0+071126 & 168 & 168  &  160913.3+264129 & 336 &  336&  232159.8+273246 & 450 &  422 \\  
042952.9+272437 & 189 & 189  &104117.1+061016 & 163 & --   &  161637.5+045932 & 461 &  369&  233040.8+110018 & 328 &  323 \\  
043103.7+203734 & 428 & --   &104244.6+120331 & 166 & 166  &  161903.6+061302 & 538 &  538&  234029.0+264156 & 469 &  422 \\ 
043337.8+290555 & 681 & 681  &105148.7+211952 & 205 & --   &  162439.0+234512 & 399 &  399&  234636.8+093045 & 262 &  --  \\ 
\hline                        
\end{tabular}                 
}                                                                                                                                                         
\end{table*}
%%%%%%%%%%%%%%%%%%%%%%%%%%%%%%%%%%%%
%%%%%%%%%%%%%%%%%%%%%%%%%%%%%%%%%%%%%%%%%%                                                                                                               

\begin{figure}[tbh]
\includegraphics*[width=8.0cm]{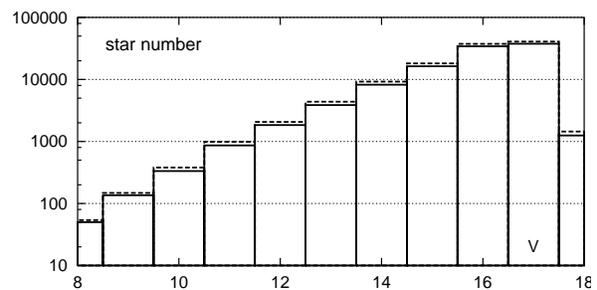} 
\caption{Distribution of the KMAC1-T (solid) and KMAC1-CU (dashed) 
star magnitudes
}
\label{figdis}
\end{figure}

The distribution of stars by  magnitude (Fig.~\ref{figdis}) 
shows that the catalogue limiting magnitude is near V=17.0~mag.
Note that beyond this limit, some of the faint
V$>$17~mag objects in the catalogue may                                                                                                                                                        
be artifacts appearing due to the  low detection 
threshold. A substantial ratio of very faint stars 
were not identified with any of the 2MASS, USNO-A2.0 or
UNSO-B1.0 objects, which could be related to either variability
of stars or  false detection. Thus, while 99.3\% of stars to 16~mag
were found in one of the major catalogues, 
this ratio drops to 77\% for V$>17$~mag stars, and for yet
fainter V$>17.5$~mag stars the ratio 
decreases to 56\%.
Nevertheless,
faint  stars were not excluded from the catalogue because
of the very low probability of false detections since each star 
was observed  in at least  two CCD scans. 
A very powerful indicator of the detection feasibility is the number
of times the star was observed.  Thus,  among V$>17.5$~mag stars
observed at least 3 times, the ratio of identifications
with external catalogues is 98.3\%.

%%%%%%%%%%%%%%%%%%%%%%%%%%%%%%%%%%%%%%%%%%

\begin{figure}[tbh]
\includegraphics*[width=8.0cm]{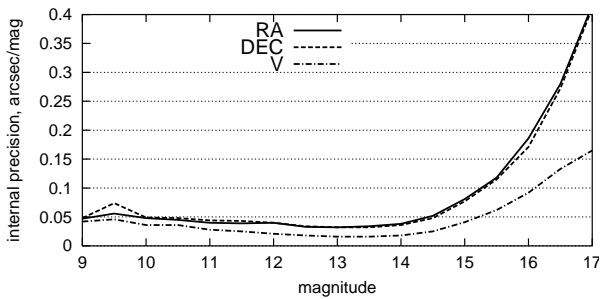} 
\caption{Internal mean accuracy of one catalogue entry
as a function of magnitude}
\label{intacc}
\end{figure}

The internal precision of the catalogue was estimated 
in a somewhat unconventional way, by comparing
CCD positions and instrumental magnitudes of stars in those nights
when they were observed. The comparison was made with  nightly 
scans transformed to the equivalent scan system (Sect.~3.2). Note
that an important feature of this transformation procedure 
is a scan to scan fitting that works as a filter which completely 
removes any systematic differences \emph {between } scans, 
leaving only random components. This is the reason why
the internal precision can be estimated in the way discussed with
no use of equatorial positions and V magnitudes computed for each
night (this data was never computed).
Results presented in Fig.~\ref{intacc} show the equal accuracy of
RA and DEC positions. 
                            
%%%%%%%%%%%%%%%%%%%%%%%%%%%%%%%%%%%%%%%%%%

\subsection{External verification of the catalogue}
\begin{figure}[tbh]
\centerline{%
\begin{tabular}{@{}c@{}}
\includegraphics*[height=120pt, width=8.0cm]{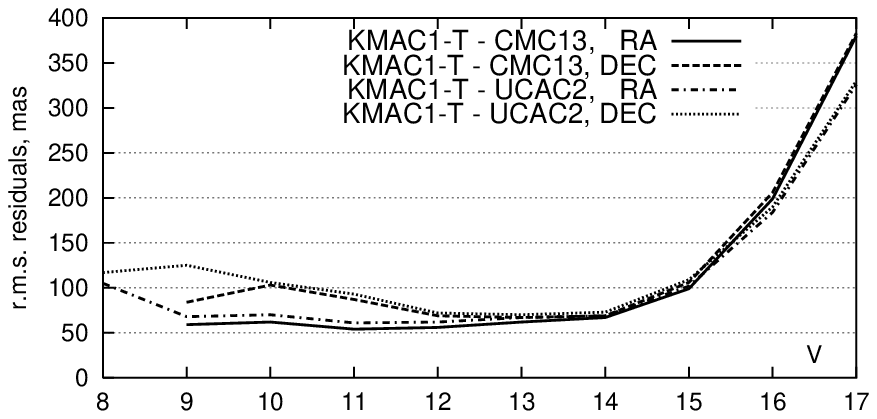} \\
\includegraphics*[height=120pt, width=8.0cm]{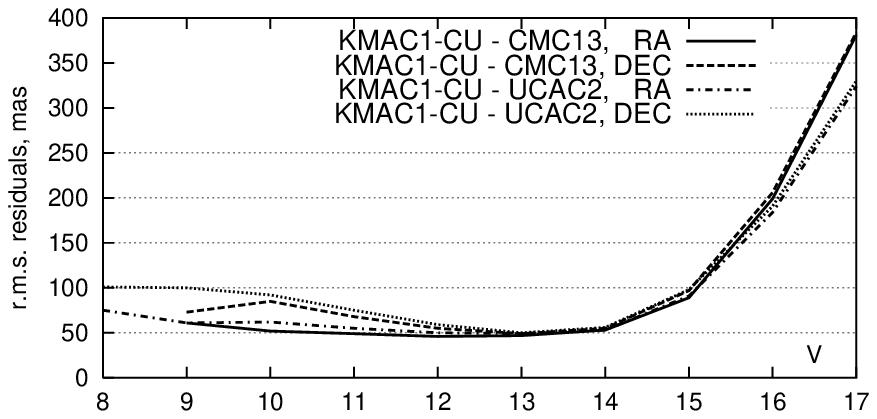} 
\end{tabular}}
\caption{ R.m.s. residuals of the KMAC1 positions
with  the CMC13 and UCAC2; upper panel -- for the version "T";
bottom -- for the version "CU"}
\label{acc}
\end{figure}

External verification of the KMAC1 positional accuracy was performed using
the CMC13 and UCAC2 which are the only all-sky sources of
present epoch positions available for faint  stars. 
We computed 
the r.m.s. differences of  KMAC1 positions 
with  the CMC13 and UCAC2 (Fig.~\ref{acc}). 
The plots  that refer to  catalogue versions "T" and "CU" are very similar.
The increase
of errors at the bright V$<$12~mag end  is caused
by oversaturation of images and affects primarily declinations.
For fainter magnitudes, the precision of RA and
DEC is the same, which is  evidence of the good efficiency
of various calibrations applied to improve declination data.

\begin{figure}[tbh]
\includegraphics*[width=8.0cm]{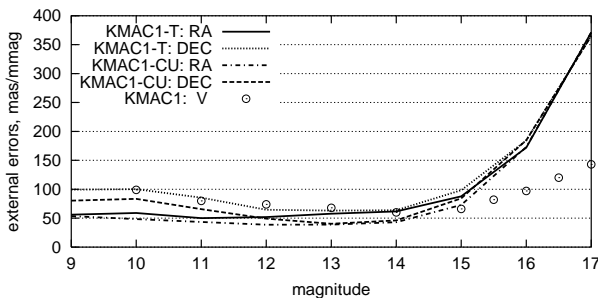} 
\caption{External errors  of the KMAC1 ("T" and "CU" versions)
positions (curves) and photometry (open circles)}
\label{ext}
\end{figure}

Plots shown in Fig.~\ref{acc} include errors of comparison
catalogues and so mark the upper limit of KMAC1 errors. 
More correct estimation of KMAC1 errors requires use of information
on the quality of the comparison catalogues.
The data on the external accuracy of the CMC13 is given by 
Evans et al. (\cite{evans}) and formal (internal) errors 
of the UCAC2 are given by Zacharias et al. (\cite{ucac}).
With  this information, 
we estimated external KMAC1 positional errors
(Fig.~\ref{ext}) separately for the "T" and "CU" catalogue versions.
The worse quality of the KMAC1-T catalogue is due to problems
with reduction to the ICRF system in short CCD scans  containing
a limited sample of Tycho2 reference stars. Note that
external and internal (Fig.~\ref{intacc}) 
errors of the KMAC1-CU are almost equal, which
indicates a very good referencing of instrumental positions to the
equatorial system.

%%%%%%%%%%%%%%%%%%%%%%%%%%%%%%%%%%%%%%%%%%
%%%%%%%%%%%%%%%%%%%%%%%%%%%%%%%%%%%%%%%%%%

We were not able to perform a direct external verification
of the KMAC1 magnitudes because of the lack of all-sky referencing available
for faint stars in the V band. Comparison  to the Valinhos
photometry (Camargo et al. \cite{camargo}) was performed for 
a 1\% subset of catalogue stars and yielded  about a 0.1~mag error
estimate (see Sect.4).   

External photometric errors (Fig.~\ref{ext})  were found considering 
the dispersion of  points in Fig.~\ref{m5}, which are the deviations 
of V-r$'$ residuals from the color calibration curve
(\ref{eq:m4}). 
No subtraction of CMC13 photometric errors was applied since
for magnitudes fainter than V=16 these errors 
were found to exceed the r.m.s.  KMAC1-CMC13 differences. 
Thus, at V=16.4~mag (16.0~mag in the r$'$ band)
the r.m.s.  KMAC1-CMC13 differences are equal to 0.12~mag while
the CMC13 external error is 0.17~mag. Probably, the quality of the CMC13 
photometry is  better than  cited.

\begin{figure}[tbh]
\centerline{%
\begin{tabular}{@{}c@{}}
\includegraphics*[height=130pt, width=8.0cm]{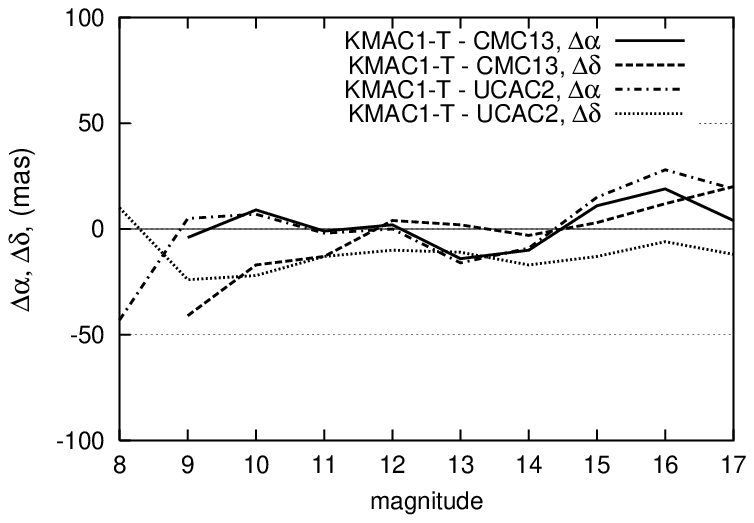} \\
\includegraphics*[height=130pt, width=8.0cm]{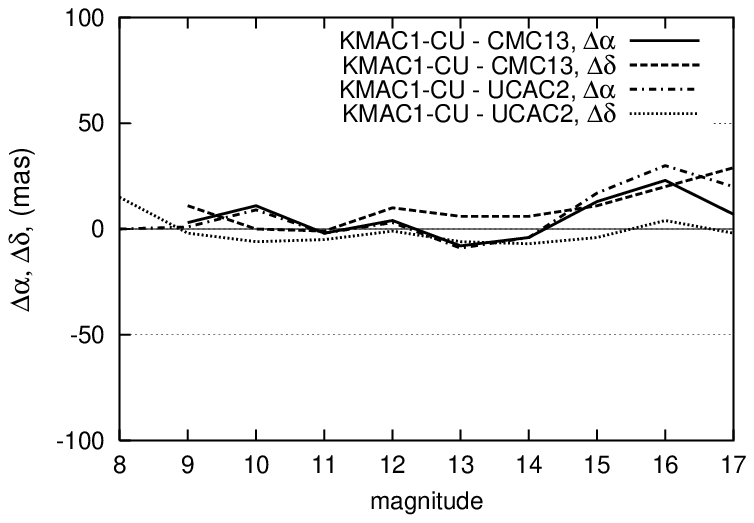} 
\end{tabular}}
\caption{Systematic differences $\Delta \alpha$ and $\Delta \delta $
of the KMAC1-T and 
KMAC1-CU positions to  those in the CMC13 and UCAC2 as a function
of magnitude}
\label{syst}
\end{figure}

Magnitude-dependent systematic errors of KMAC1-CU and KMAC1-T
do not exceed $\pm 20$~mas and $\pm 40$~mas respectively
(Fig.~\ref{syst}.). No clear dependency on magnitude is  seen
except for a positive hump of  plots at V$\approx$16~mag, and
a negative downtrend for the version KMAC1-T in DEC at the
bright V$<$11 end.

\begin{figure}[tbh]
\begin{tabular}{@{}c@{}}
\includegraphics*[bb =  32 465 546 818 height=120pt, width=8.0cm]{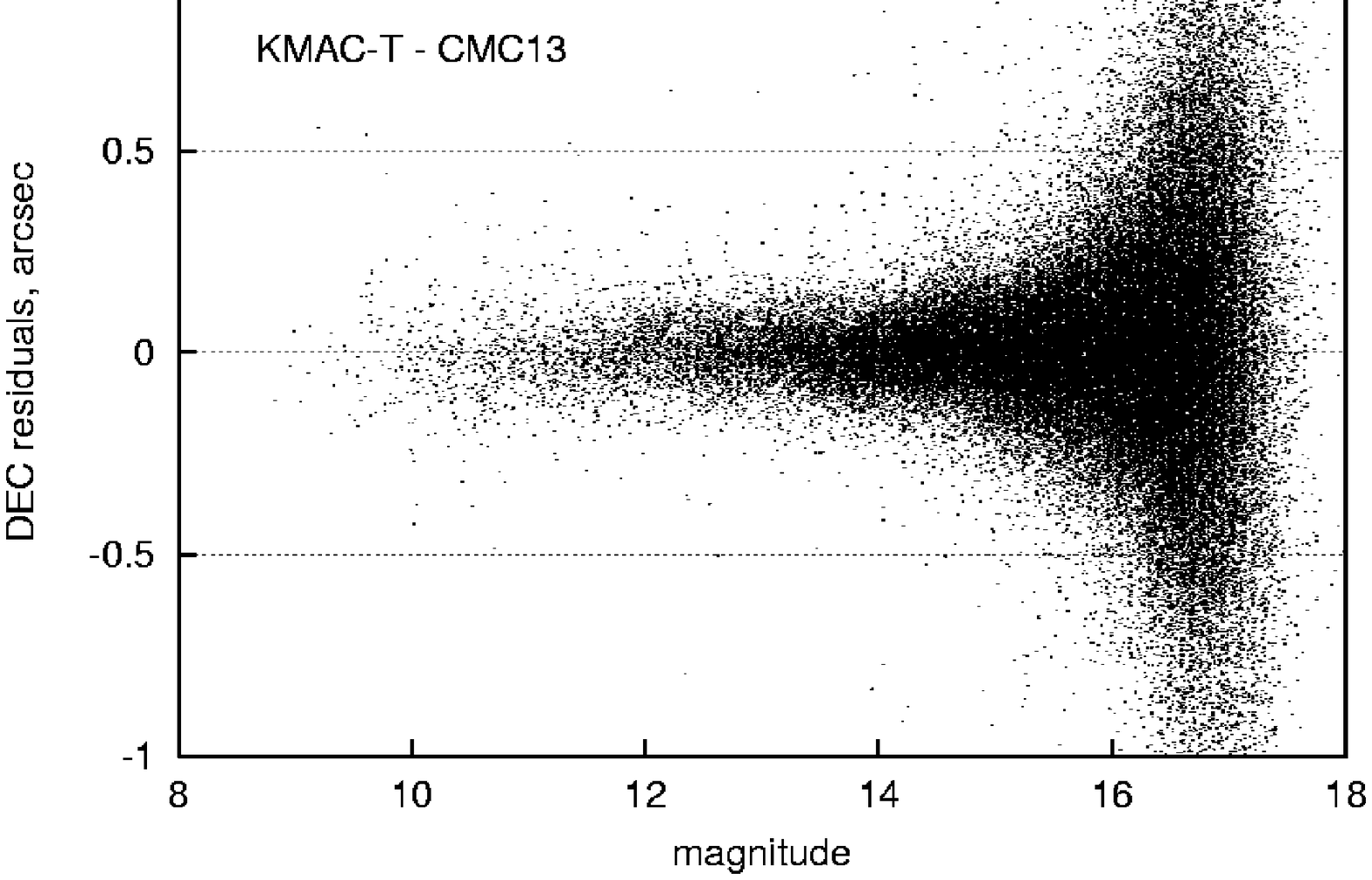} \\
\includegraphics*[bb = 44 467 566 825 height=120pt, width=8.0cm]{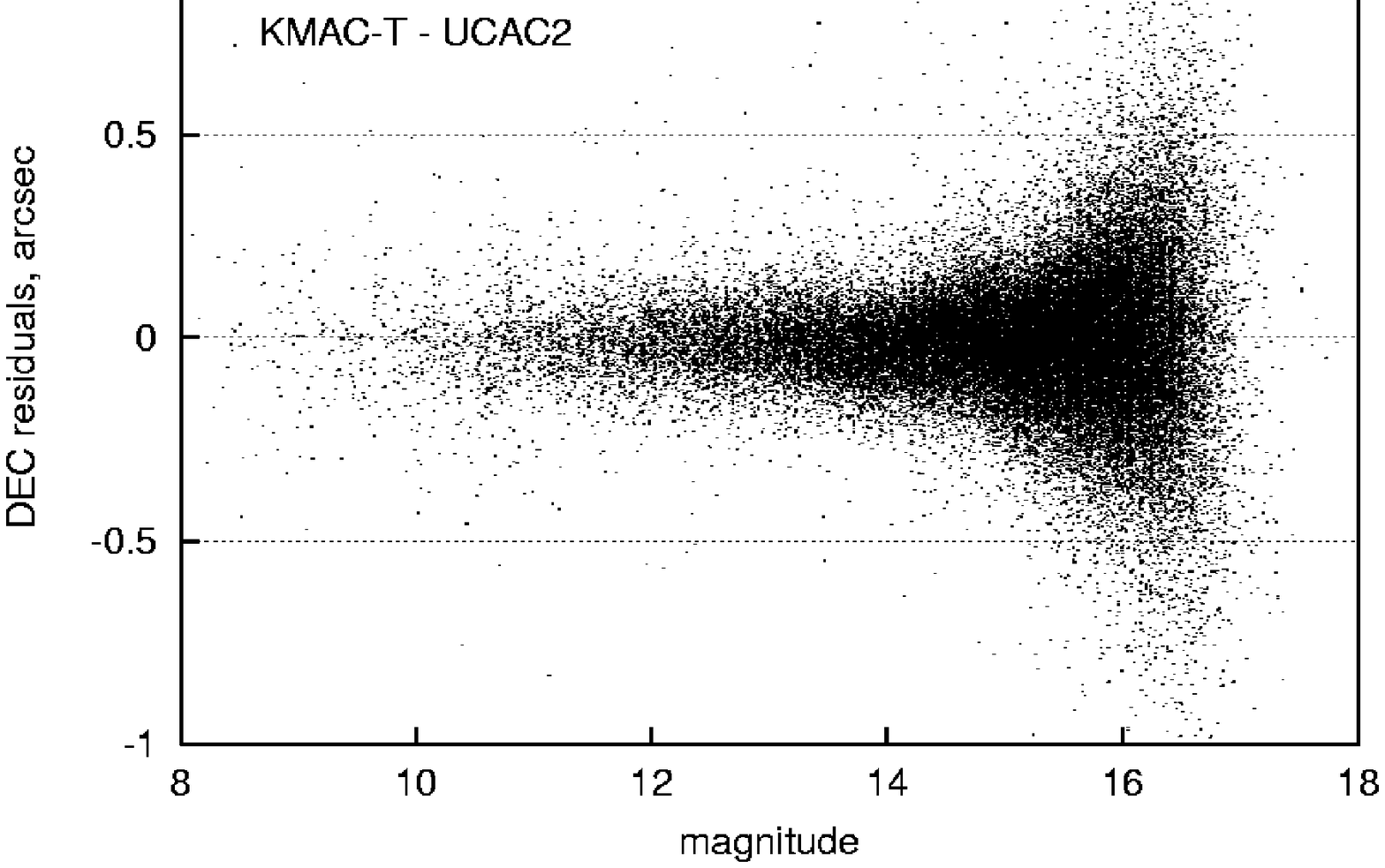} 
\end{tabular}
\caption{Individual  differences of declinations in the KMAC1-T and
in CMC13 and UCAC2} 
\label{indiv}
\end{figure}

%%%%%%%%%%%%%%%%%%%%%%%%%%
Individual  differences between  the KMAC1-T  star positions and
positions in the comparison catalogues CMC13 and UCAC2 are shown 
in Fig.~\ref{indiv}. We present the worst comparison in DEC
and the "T" catalogue version; 
slightly better plots can be obtained for RA and  
the KMAC1-CU catalogue version. No systematic trend of 
individual differences
in magnitude is observed.

%%%%%%%%%%%%%%%%%%%%%%%%%%

\section{Conclusion}
The aim of this work was to obtain a catalogue of faint stars
in  sky areas with ICRF objects whose declinations are
optimal for observations with the MAC. The catalogue contains positions
of faint V$<$17~mag objects referred to the optical Hipparcos-Tycho
reference frame and thus presents an extension of the ICRF to the
optical domain.

The catalogue described in this Paper is the first catalogue obtained
with the Kyiv meridian axial circle after it was refurbished with a CCD
camera. Realization of this project  involved  development
of  special software for image processing, astrometric calibration
for instrumental errors etc. A quite unexpected finding was that the measured
data (especially the DEC component) is strongly 
affected by systematic errors even when star images have a relatively good 
shape. A solution to this problem was found in extensive use of external
astrometric catalogues for calibrations.

Another difficulty arose from  underestimation of the Tycho2 
errors at the present epoch and from the inhomogeneous sky
distribution of the catalogue stars. 
As a result, scan lengths  appeared to be too short
to allow rigorous reduction to the ICRF
and forced us to use other catalogues (CMC13 and UCAC2) for referencing. 
The use of the Tycho2
catalogue for astrometric work in small fields of about 0.5x0.5$^{\circ}$
or less is thus problematic, and feasible only in some sky areas.